\documentclass[12pt]{article}
\usepackage{amsmath,amsfonts}
\usepackage{amscd,amsthm}
\usepackage{geometry}
\usepackage{graphicx}

\newtheorem{theorem}{Theorem}

\newtheorem{corollary}[theorem]{Corollary}

\newtheorem{proposition}[theorem]{Proposition}

\newcommand{\documentdate}{23 September 2010}
\renewenvironment{proof}[1][Proof]{\noindent\textbf{#1.} }{\ \rule{0.5em}{0.5em}}
\begin{document}

\begin{titlepage}

\includegraphics[height=3.5cm]{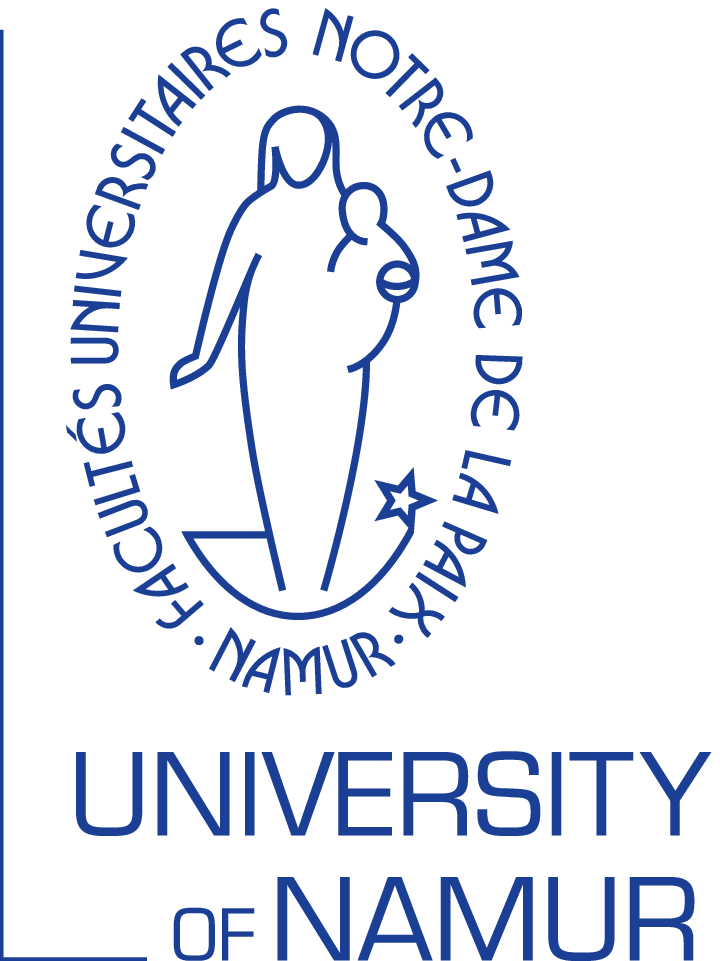}

\vspace*{2cm}
\hspace*{1.3cm}
\fbox{\rule[-3cm]{0cm}{6cm}\begin{minipage}[c]{12cm}
\begin{center}
{\Large Adaptive Expectations, Confirmatory Bias, and Informational
Efficiency}\\
\mbox{}\\
by Gani Aldashev, Timoteo Carletti and Simone Righi\\
\mbox{}\\
Report naXys-02-2010 \hspace*{20mm} \documentdate 
\end{center}
\end{minipage}
}

\vspace{3cm}
\begin{center}
\includegraphics[height=3.5cm]{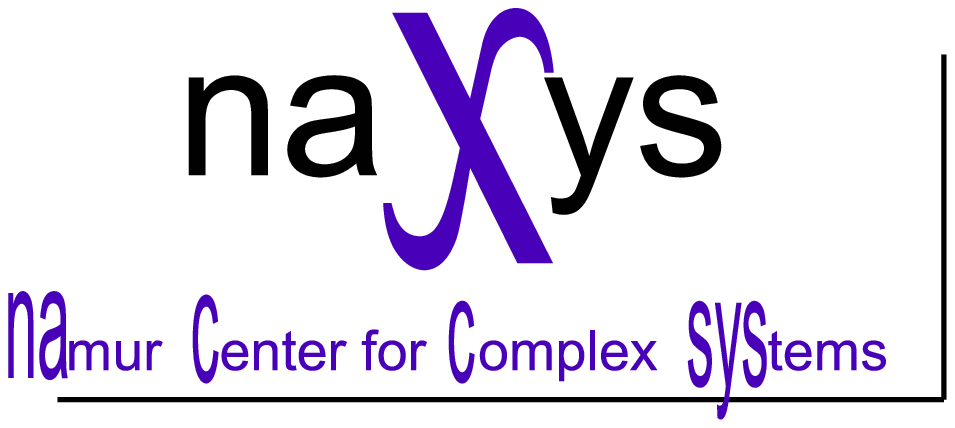}

\vspace{1.5cm}
{\Large \bf Namur Center for Complex Systems}

{\large
University of Namur\\
8, rempart de la vierge, B5000 Namur (Belgium)\\*[2ex]
{\tt http://www.naxys.be}}

\end{center}

\end{titlepage}

\newpage

\title{{\huge Adaptive Expectations, Confirmatory Bias, and Informational
Efficiency}\thanks{%
The authors thank the National Bank of Belgium for financial support.}}
\author{Gani Aldashev\thanks{%
Corresponding author. Department of Economics and CRED, University of Namur
(FUNDP). Mailing address: Department of Economics, 8 Rempart de la Vierge,
5000 Namur, Belgium. Email: gani.aldashev@fundp.ac.be.} \and Timoteo Carletti%
\thanks{%
NaXys - Namur Center for Complex Systems, University of Namur (FUNDP).
Mailing address: 8 Rempart de la Vierge, 5000 Namur, Belgium. Email:
timoteo.carletti@fundp.ac.be.} \and Simone Righi\thanks{%
Department of Economics, University of Namur (FUNDP). Mailing address:
Department of Economics, 8 Rempart de la Vierge, 5000 Namur, Belgium. Email:
simone.righi@fundp.ac.be.}}
\maketitle

\begin{abstract}
We study the informational efficiency of a market with a single traded
asset. The price initially differs from the fundamental value, about which
the agents have noisy private information (which is, on average, correct). A
fraction of traders revise their price expectations in each period. The
price at which the asset is traded is public information. The agents'
expectations have an adaptive component and a social-interactions component
with confirmatory bias. We show that, taken separately, each of the
deviations from rationality worsen the information efficiency of the market.
However, when the two biases are combined, the degree of informational
inefficiency of the market (measured as the deviation of the long-run market
price from the fundamental value of the asset) can be non-monotonic both in
the weight of the adaptive component and in the degree of the confirmatory
bias. For some ranges of parameters, two biases tend to mitigate each
other's effect, thus increasing the informational efficiency.\bigskip
\bigskip


\textit{JEL\ codes}: G14, D82, D84.
\end{abstract}

\newpage

\setlength{\baselineskip}{20pt}

\section{Introduction}

In most economic interactions, individuals possess only partial information
about the value of exchanged objects. For instance, when a firm "goes
public", i.e. launches an initial public offering of its shares, no
financial market participant has the complete information concerning the
future value of the profit stream that the firm would generate. The
fundamental question, going back to Hayek (1945), is then: To which extent
the market can serve as the aggregator of this dispersed information? In
other words, when is the market informationally efficient, i.e. that the
market price converges to the value that would obtain if all market
participants were to have full information about the fundamental value of
the asset exchanged?

Most of the studies that address this question are based on the assumption
that individual market participants are rational. Under full rationality,
the seminal results on the informational efficiency of centralized markets
have been established by Grossman (1976), Wilson (1977), Milgrom (1981),
and, for decentralized markets, by Wolinsky (1990), Blouin and Serrano
(2001), and Duffie and Manso (2007).

However, research in experimental economics and behavioral finance indicates
that traders do not behave in the way consistent with the full-rationality
assumption. For instance, Haruvy et al. (2007) find that traders have
adaptive expectations, i.e. they give more importance to the past realized
price of the asset than the fully-rational agent would. Along a different
dimension, Rabin and Schrag (1999) discuss the evidence that individuals
suffer from the so-called confirmatory (or confirmation) bias: they tend to
discard the new information that substantially differs from their priors.
Understanding whether (and under which conditions) the financial markets are
informationally efficient when agents do not behave fully rationally remains
an open question.

In this paper, we study the informational efficiency of a market with a
single traded asset. The price initially differs from the fundamental value,
about which the agents have noisy private information (which is, on average,
correct). A fraction of traders revise their price expectations in each
period. The price at which the asset is traded is public information. The
agents' expectations have an adaptive component and a social-interactions
component with confirmatory bias.

We show that, taken separately, each of the deviations from rationality
worsen the information efficiency of the market. However, when the two
biases are combined, the degree of informational inefficiency of the market
(measured as the deviation of the long-run market price from the fundamental
value of the asset) can be non-monotonic both in the weight of the adaptive
component and in the degree of the confirmatory bias. For some ranges of
parameters, two biases tend to mitigate each other's effect, thus increasing
the informational efficiency.

The paper is structured as follows. Section 2 presents the setup of the
model. Section 3 derives analytical results for each bias taken separately.
In Section 4, we present the simulation results when two biases are
combined. Finally, Section 5 discusses the implication of our results and
suggests some future avenues for research.

\section{The model}

Consider a market with $N$ participants, each endowed with an initial level
of wealth equal to $W_{0}>0$. The amount $L_{0}\in (0,W_{0}]$ is in liquid
form. Time is discrete (e.g. to mimic the daily opening and closure of a
financial market), denoted with $t=0,1,...$. Market participants trade a
single asset, whose price in period $t$ we denote with $P_{t}$. This price
is public information. Prices are normalized in such a way that they belong
to the interval $[0,1]$.

At the beginning of each period $t$, every agent $i$ can place an order to
buy or short sell $1$ unit of the asset, on the basis of her expectation
about the price for period $t$, denoted with $P_{t}^{e,i}$. Placing an order
implies a fixed, small but positive \emph{transaction cost} $c$, i.e. $%
0<c\ll 1$. At the end of the period, each agent $i$ learns the price $P_{t}$
at which the trade is settled (as explained below).

The agent $i$ then constructs her price expectation for the next period and
decides to participate in the trading in period $t+1$ according to the \emph{%
expected next-period gain}, i.e. if%
\begin{equation}
\left\vert P_{t+1}^{e,i}-P_{t}\right\vert -c>0\,.  \label{eq:expnexpergain}
\end{equation}

Moreover, she participates as a \emph{buyer} if her price expectation for
the next period exceeds the current price, i.e.%
\begin{equation}
P_{t+1}^{e,i}>P_{t}\,,  \label{eq:buyer}
\end{equation}%
or as a \emph{seller} if, on the contrary,%
\begin{equation}
P_{t+1}^{e,i}<P_{t}\,.  \label{eq:seller}
\end{equation}

The way in which agents form their next-period price expectations differs
from the standard rational-expectation benchmark in the following way. First
deviation is the fact that agents give weight to the past public prices,
i.e. they have (partially) adaptive expectations. Secondly, they can
influence each other's expectations via social interactions with
confirmatory bias.

Formally, suppose that in every period a fraction, $\gamma \in \lbrack 0,1]$%
, of the agents makes a revision of their price expectations. An agent
revises her price expectation by analyzing the past price of the asset and
by randomly encountering some other agent (at zero cost), and possibly
exchanging her own price expectation with this partner. In these encounters,
the agents have a \emph{confirmatory bias}, i.e. each agent tends to ignore
the information coming from the other agent if it differs too much from her
own. If, on the contrary, this difference is not too large , i.e. smaller
than some fixed threshold, which we denote with $\sigma $, then the agent
incorporates this information into her price expectation. The remaining $%
(1-\gamma )N$ agents do not revise their expectations in the current period.

Summarizing, the expectation formation process of agent $i$ meeting agent $j$
is: 
\begin{equation}
P_{t+1}^{e,i}=\alpha P_{t}+(1-\alpha )%
\begin{cases}
P_{t}^{e,i} & \text{if }\left\vert P_{t}^{e,i}-P_{t}^{e,j}\right\vert \geq
\sigma \\ 
{\displaystyle\frac{P_{t}^{e,i}+P_{t}^{e,j}}{2}} & \text{otherwise}%
\end{cases}%
\,,  \label{eq:priceupdate}
\end{equation}%
and it is analogous for $P_{t+1}^{e,j}$. Here, $\alpha $ measures the
relative weight of the past price. If $\alpha =1$, the agents have purely
adaptive expectations (and social interactions play no role). If $\alpha =0$
and $\sigma =1$ the agents (that revise their expectations) completely
disregards the past and fully integrate all the information coming from the
social interactions.

Our objective is to analyze the price formation under the different values
of the parameters $\alpha $, $\sigma $, and $\gamma $.

Concerning the market microstructure, we assume that the market is
centralized, with a simple price response to excess demand. In other words,
the market mechanism is similar to the Walrasian auctioneer. More precisely,
the price formation mechanism functions as follows:

\begin{enumerate}
\item \label{item:pstar} There exists a hypothetical price at period $t+1$
that would (approximately) equate the number of buy-orders and sell-orders.
Let us denote it with $P_{t+1}^{\ast }$. From~\eqref{eq:buyer} and~%
\eqref{eq:seller}, $P_{t+1}^{\ast }$ is the solution of the equation: 
\begin{equation*}
n_{B}(x)=n_{S}(x)\,,
\end{equation*}%
where $n_{B}(x)$ and $n_{S}(x)$ are the numbers of buyers and sellers at
price $x$. Whenever there are several solutions to this equation, $%
P_{t+1}^{\ast }$ denotes the average of the values that solve the equation.

\item Out of equilibrium, the price adjustment depends on the size of the
excess demand or excess supply relative to the size of the population; in
other words, denoting $\beta (x)=\left\vert n_{B}(x)-n_{S}(x)\right\vert /N$%
, the price adjustment process is: 
\begin{equation}
P_{t+1}=\beta (P_{t})P_{t+1}^{\ast }+(1-\beta (P_{t}))P_{t}\,.
\label{eq:pricetp1}
\end{equation}%
Thus, the deviation from the equilibrium does not disappear instantly.
However, the price moves in the direction that eliminates the excess demand
or supply, and, moreover, the speed of adjustment depends on the size of
disequilibrium (relative to the size of the population).\footnote{%
We avoid the shortcoming of assuming a constant $\beta (P_{t})$. As
discussed by LeBaron (2001), if $\beta (P_{t})$ is assumed to be constant,
the behavior of the simulated market is extremely sensitive to the value of $%
\beta $, which makes it difficult to interpret the results.}

\item Given that each agent that participates in the market in period $t$
places an order for one unit of the asset, the number of exchanges that
occurs is $\min \{n_{B}(P_{t}),n_{S}(P_{t})\}$. Then, each seller $i$
updates her wealth by $W_{t+1}^{i}=W_{t}^{i}+P_{t}-P_{t+1}-c$, and her
liquidity by $L_{t+1}^{i}=L_{t}^{i}+P_{t}-c$. Similarly, for a buyer $j$, we
have $W_{t+1}^{j}=W_{t}^{j}-P_{t}+P_{t+1}-c$ and $%
L_{t+1}^{j}=L_{t}^{j}-P_{t}-c$.

\item If an agent's liquidity dries up to zero, then she leaves the market.
At her place, at the beginning of the next period enters a new agent with
wealth $W_{0}$, liquidity $L_{0}$, and the next-period price expectation
randomly drawn from the $[0,1]$ interval.
\end{enumerate}

In this setting, consider an initial public offering (IPO) of the asset. At
time $t=0$, the asset gets introduced in the market at some price $P_{0}$.
Let's also suppose that, \textit{on average}, the agents have the unbiased
information about its fundamental value. In particular, let's suppose that
the initial price expectations of the agents is a uniform distribution in
the $[0,1]$ interval, i.e. the fundamental value of the asset is $\frac{1}{2}
$. However, the initial price $P_{0}$ differs from the fundamental value.
The questions that we pose are:

\begin{itemize}
\item Does the market price $P_{t}$ converge to the fundamental value of the
asset?

\item If not, how large is the deviation of the long-run price $P_{t}$ as $%
t\rightarrow \infty $, from the fundamental value?

\item How does this deviation depend on the weight of history $\alpha $
(i.e. the "adaptiveness" of agents' expectations), the prominence of the
confirmatory bias of the traders $\sigma $, and the frequency with which
agents adjust their expectations, $\gamma $?
\end{itemize}

\section{Analytical results}

We can characterize analytically the answers to the above questions for some
of the values of the parameters. This requires a further assumption that the
number of market participants ($N$) and every agent's initial wealth and
liquidity ($W_{0}$ and $L_{0}$) are sufficiently large.

\subsection{Purely adaptive expectations}

Consider first the case where agents discard the social interactions and
consider only the past price. In other words, $\alpha =1$ in~%
\eqref{eq:priceupdate}. We analyze separately two sub-cases: (i) all agents
revise their expectations in every period, i.e. $\gamma =1$; and (ii) only a
fraction of agents revise their expectations in every period, i.e. $\gamma
<1 $.

(i) In the case $\gamma =1$, Eq.~\eqref{eq:priceupdate} simply reduces to:%
\begin{equation*}
{P}_{t+1}^{e,i}=P_{t}\,\ \text{for any }i\text{ and }t.
\end{equation*}%
However, this is also the hypothetical price that equates buyers and
sellers, i.e. $P_{t+1}^{\ast }=P_{t}$, and thus $\beta (P_{t})=0$. Finally,
from~\eqref{eq:pricetp1} we get $P_{t+1}=P_{t}$. This means that the market
price doesn't evolve: $P_{t}=P_{0}$ in every period. Intuitively, if all
agents revise their expectations in every period and have purely adaptive
expectations, once the initial price $P_{0}$ is announced, every agent
immediately revises her next-period expectation, substituting it with $P_{0}$%
. Given that every agents does so, no agents is interested in trading, and
the price does not evolve.

(ii) Next, consider the case $\gamma <1$, with $\gamma N$ being sufficiently
large. Without loss of generality, suppose that $P_{0}>1/2$. We prove that
the market reaches the long-run equilibrium, after a few periods, with the
long-run market price deviating from the initial price by a value smaller
than $c(1-\gamma )$.

We need the following preliminary result.

\begin{proposition}
\label{prop:pstarcomput} Consider a population of agents divided into two
groups: agents in the first group, whose size is $N_{1}$, have expectations
uniformly distributed in $[0,1]$, and agents in the second group, whose size
is $N_{2}>>N_{1}$, all have the price expectation equal to some fixed $\hat{P%
}\in (0,1)$. Then, the price $P^{\ast }$ defined at point~\ref{item:pstar}
is given by $P^{\ast }=\hat{P}-c$ if $\hat{P}>1/2$ and $P^{\ast }=\hat{P}+c$
if $\hat{P}<1/2$. If $\hat{P}=1/2$, then $P^{\ast }=\hat{P}$.
\end{proposition}

\proof We consider only the case $\hat{P}>1/2$ (the proof for the case $\hat{%
P}<1/2$ is analogous, while the case $\hat{P}=1/2$ is trivial). Define the
functions $\theta _{c}(x,P)$ 
\begin{equation*}
\theta _{c}(x,P)=%
\begin{cases}
1 & \text{if $x>P+c$} \\ 
0 & \text{otherwise}%
\end{cases}%
\,,
\end{equation*}%
and $\eta _{c}(x,P)=1-\theta _{-c}(x,P)$. Then for a sufficiently large $%
N_{2}$, the numbers of sellers and buyers at price $x\in \lbrack 0,1]$ are
respectively given by:%
\begin{eqnarray}
n_{S}(x) &=&(x-c)N_{1}+N_{2}\theta _{c}(x,\hat{P})\quad \text{and}
\label{eq:buysell1} \\
n_{B}(x) &=&(1-x-c)N_{1}+N_{2}\eta _{c}(x,\hat{P}).  \notag
\end{eqnarray}

This follows from the trading protocol, given that for a price sufficiently
close to $\hat{P}$ (i.e. with a deviation less than $c$), only the first
group of agents participates in the trading, and that the expectations are
uniformly distributed in the first group. On the other hand, if $x<\hat{P}-c$
(or $x>\hat{P}+c$), the second group also participates in the trading as
buyers (sellers).

Then, the difference in the number of buyers and sellers is%
\begin{equation}
\Delta (x)=(1-2x)N_{1}+N_{2}\left( \eta _{c}(x,\hat{P})-\theta _{c}(x,\hat{P}%
)\right) ,  \label{eq:buysell2}
\end{equation}%
and, therefore, $P^{\ast }$ becomes the price at which the sign of $\Delta
(x)$ changes (or the average of these values, if more than one exist). We
can then easily prove that 
\begin{equation}
\Delta _{-}=\lim_{x\rightarrow (\hat{P}-c)^{-}}\Delta (x)=(1-2\hat{P}%
+2c)N_{1}+N_{2}>0.  \label{eq:buysell3}
\end{equation}%
Finally, using the assumption $N_{2}>>N_{1}$, we get $\Delta _{-}>0>\Delta
(x)$ for all $x>\hat{P}-c$. This implies that $P^{\ast }=\hat{P}-c$. 
\endproof

This proposition has the following

\begin{corollary}
Suppose the assumptions of Proposition 1 hold. If a third group of agents
(of arbitrary size) with price expectation $\tilde{P}$, such that $|\tilde{P}%
-\hat{P}|<c$, joins the market, then $P^{\ast }$ does not change.
\end{corollary}

We can now analyze the market dynamics under the assumptions $\alpha =1$ and 
$\gamma <1$ with $\gamma N$ large.

During the first period $N_{1}=(1-\gamma )N$ agents do not revise their
expectations. These expectations are uniformly distributed in $[0,1]$.
Contrarily, $N_{2}=\gamma N$ agents revise their expectations, which now
becomes the IPO price $P_{0}$ (i.e. $P_{1}^{e,i}=P_{0}$). Proposition 1
ensures that $P_{1}^{\ast }=P_{0}-c$. Moreover, the size of market
disequilibrium is small: using the definition, we get $\beta
(P_{0})=(2P_{0}-1)(1-\gamma )$. Finally, the end-of-period 1 price $P_{1}$
will be%
\begin{equation}
P_{1}=\beta (P_{0})(P_{0}-c)+(1-\beta (P_{0}))P_{0}=P_{0}-\beta (P_{0})c.
\label{eq:buysellnewpri}
\end{equation}%
Note that this price is $c$--close to $P_{0}$, given that $\beta (P_{0})$ is
small.

During the next period, $\gamma N$ agents revise their expectations, while $%
(1-\gamma )N$ agents do not revise. Then, on average, we have that the
second group contains $N_{2}=\gamma (1-\gamma )N$ agents, for which $%
P_{2}^{e,i}=P_{1}^{e,i}=P_{0}$, the first group contains $N_{1}=(1-\gamma
)^{2}$ agents (who do not revise their initial expectations), and, moreover,
there exists a third group of agents, of size $N_{3}=\gamma (2-\gamma )$,
for whom $P_{2}^{e,i}=P_{1}=P_{0}-\beta (P_{0})c$. We can then apply
Corollary 2 and conclude that $P_{2}^{\ast }=P_{0}-c$. Computing the
next-period market disequilibrium $\beta (P_{1})$, we can easily observe
that $\beta (P_{1})\sim (1-\gamma )^{2}$. Therefore, the next-period price $%
P_{2}$ will be%
\begin{equation}
P_{2}=\beta (P_{1})(P_{0}-c)+(1-\beta (P_{1}))P_{1}\sim P_{0}-\beta (P_{0})c+%
\mathcal{O}(1-\gamma )^{2}\sim P_{0}-c(1-\gamma )+\mathcal{O}(1-\gamma
)^{2}\,.  \label{eq:buysellnewpri2}
\end{equation}

Thus, the market price varies as long as there exist agents that have not
yet revised their initial expectations. However, the market price does not
move too far from $P_{0}$. Assuming the extreme-case scenario where in every
period the same $(1-\gamma )$ agents happen to be the ones that do not
revise their expectations, the number of periods that pass before the market
price converges to its steady-state value equals $-\log N/\log (1-\gamma )$.

Numerical simulations presented in Figure 1 confirm our theoretical findings.

\begin{center}
\textbf{[Insert Figure 1 about here]}
\end{center}

\subsection{Social interactions}

Next, consider the setting where the agents' expectations have no adaptive
component (i.e. $\alpha =0$), and the agents that revise their expectations
rely on the social interactions with other agents. Then, the relevant
parameter is the extent of the confirmatory bias ($\sigma $) that the agents
have. We derive analytical results for the the cases of the extreme form of
confirmatory bias ($\sigma <<1$) and for that of virtually no bias ($\sigma
\sim 1$).

\subsubsection{Social interactions: large confirmatory bias}

Consider the extreme form of confirmatory bias, i.e. that whenever two
agents meet, neither of them adjusts her price expectation, no matter how
close their past-period expectations are. We will prove that the market is
fully informationally efficient in the long-run, but convergence to this
efficient outcome takes arbitrary long time.

In every period, $\gamma N$ agents engage in social interactions (without
influencing each other's price expectations). This implies that no agent
revises her price expectation. Therefore, the mean price expectation (which
we denote with $\widetilde{P}^{e}$) does not change either; namely, from~%
\eqref{eq:priceupdate},%
\begin{equation*}
\widetilde{P}_{t+1}^{e}=\widetilde{P}_{t}^{e}.
\end{equation*}%
Under the assumption that the initial price expectations are uniformly
distributed, we obtain $P_{t+1}^{\ast }=\widetilde{P}_{t+1}^{e}=1/2$. The
market disequilibrium is thus given by%
\begin{equation}
\beta (P_{t})=|1-2P_{t}|.  \label{eq:soc1}
\end{equation}

Therefore, the market price evolves according to the equation%
\begin{equation}
P_{t+1}=\frac{|1-2P_{t}|}{2}+\left( 1-|1-2P_{t}|\right) P_{t}=P_{t}+\frac{%
|1-2P_{t}|(1-2P_{t})}{2}.  \label{eq:socpriceevol}
\end{equation}

Let us define the mapping%
\begin{equation}
f(P)=%
\begin{cases}
P-\frac{(2P-1)^{2}}{2} & \text{if $P\geq 1/2$} \\ 
P+\frac{(2P-1)^{2}}{2} & \text{if $P<1/2$}%
\end{cases}%
.  \label{eq:socpricevolmap}
\end{equation}

The evolution of the market price is determined by the dynamic system%
\begin{equation}
P_{t+1}=f(P_{t}).  \label{eq:socdynsys}
\end{equation}%
This mapping has a unique fixed point at $P=1/2$. Moreover, this is an
attractor, whose strength decreases the closer we are to the fixed point: $%
P_{t}-1/2\sim a/t$.

Finally, note that if $P_{0}=1/2$, then~\eqref{eq:socpriceevol} implies that 
$P_{t}=1/2$ for all $t$.

\begin{center}
\textbf{[Insert Figure 2 about here]}
\end{center}

\subsubsection{Social interactions: small confirmatory bias}

Let us now consider the opposite extreme, i.e. the agent that uses for
updating her expectation that of her partners in social interactions even
when such expectations diverge radically from hers. Assume for the moment
that all agents revise their expectations in every period ($\gamma =1$). We
will prove that in this case the market is fully informationally efficient
in the long-run, and that convergence to this efficient outcome occurs
within in a finite number of periods (that essentially depends on the
transaction cost $c$).

Given that the expectation-revision rule~\eqref{eq:priceupdate} with $\alpha
=0$ and $\gamma =1$ preserves the mean price expectation, we trivially get%
\begin{equation*}
\widetilde{P}_{t+1}^{e}=\widetilde{P}_{t}^{e}=1/2.
\end{equation*}

This follows from the fact that the initial price expectations are uniformly
distributed in $[0,1]$, hence with average value $1/2$, which also equals
the hypothetical Walrasian-auctioneer price $P_{t}^{\ast }=1/2$. Moreover,
the equation~\eqref{eq:priceupdate} implies that price expectations follow
the Deffuant dynamic (Deffuant et al. 2000, Weisbuch et al. 2002). In other
words, the dispersion of price expectations, denoted with $\Delta _{P_{e}}$,
shrinks to zero according to (see the left panel of Figure~3):%
\begin{equation*}
\Delta _{P_{e}}(t)\sim \frac{1}{2^{t/2}}.
\end{equation*}

\begin{center}
\textbf{[Insert Figure 3 about here]}
\end{center}

Because of the transaction cost is positive, the market activity stops once
all the expectations fall inside the interval whose width is smaller than $2c
$. This happens after a time $\widehat{T}\sim -2\left( 1+\log _{2}c\right) $.

Let us assume now that the price expectations have a dispersion large
enough, so that the market activity still does not stop. Then, we can easily
compute the market disequilibrium $\beta (P_{t})$: 
\begin{equation*}
\beta (P_{t})=2^{t/2}|1-2P_{t}|,
\end{equation*}%
which implies that the next-period price is given by:%
\begin{equation}
P_{t+1}=P_{t}+2^{t/2-1}|1-2P_{t}|(1-2P_{t}).  \label{eq:socpriceevols1}
\end{equation}

Let us introduce the auxiliary variable $x$, defined as $%
P_{t}-1/2=x_{t}/2^{t/2}$. This allows us to fully describe the market price
evolution with the dynamic system given by the function $g(x)$:%
\begin{equation}
g(x)=%
\begin{cases}
\sqrt{2}x-2\sqrt{2}x^{2} & \text{if $x\geq 0$} \\ 
\sqrt{2}x+2\sqrt{2}x^{2} & \text{if $x<0$}%
\end{cases}%
.  \label{eq:socpricevolmaps1}
\end{equation}%
This mapping has three fixed points: $x=0$ (unstable), and $x=\pm (2-\sqrt{2}%
)/4$, that are stable.

We can thus conclude that, as long as the market activity continues, the
market price converges to $1/2$, given that in all cases $P_{t}=\frac{x_{t}}{%
2^{t/2}}+\frac{1}{2}\rightarrow \frac{1}{2}$. These findings are supported
by numerical simulations, whose results we report in Figure~4.

\begin{center}
\textbf{[Insert Figure 4 about here]}
\end{center}

In a similar fashion, we can study the case $\gamma \neq 1$. In this case,
in every period $\gamma N$ agents revise their expectations and, because
they have a very small confirmatory bias (i.e. $\sigma \sim 1$), they
influence each other's expectations. Then, overall there is a tendency for
the expectations to converges (because of the process driven by Eq. %
\eqref{eq:priceupdate} with $\alpha =0$). We should keep in mind, however,
that in every period $(1-\gamma )N$ agents do not revise their expectations.

Zero weight given to the past prices in forming the next-period expectation (%
$\alpha =0$) implies that the mean price expectation does not change, $%
\widetilde{P}_{t+1}^{e}=\widetilde{P}_{t}^{e}$. Hence $\widetilde{P}%
_{t}^{e}=1/2$ and $P_{t}^{\ast }=1/2$ for all $t$. Moreover the expectations
are distributed in an interval whose width (denoted with $\Delta _{P_{e}}(t)$%
) shrinks to zero, but slower than in the case $\gamma =1$. The simulations
presented in the right panel of Figure 3, allow us to see that this
narrowing of expectation dispersion follows approximately the law%
\begin{equation*}
\Delta _{P_{e}}(t)\sim \frac{1}{2^{q_{\gamma }t}},
\end{equation*}%
where $q_{\gamma }=a\gamma +b$, $a=0.61\pm 0.02$ and $b=-0.13\pm 0.01$,
independent of $\sigma $.

Thus, the price disequilibrium can be estimated as:%
\begin{equation*}
\beta (P_{t})\sim 2^{q_{\gamma }t}|1-2P_{t}|,
\end{equation*}%
which implies the following price dynamics:%
\begin{equation*}
P_{t+1}=P_{t}+2^{q_{\gamma }t-1}|1-2P_{t}|(1-2P_{t}).
\end{equation*}%
Introducing a new variable $y_{t}$ such that $P_{t}=1/2+y_{t}/2^{q_{\gamma
}t}$, we obtain the following difference equation for the evolution of $y_{t}
$:%
\begin{equation*}
y_{t+1}=2^{q_{\gamma }}y_{t}-2^{q_{\gamma }+1}|y_{t}|y_{y}\,.
\end{equation*}

This mapping has three fixed points, $y=0$ (unstable), and $y=\pm
(1-2^{q_{\gamma }})/2^{q_{\gamma }+1}$ (stable). We can finally conclude
that, similar to the results above, the market price converges to $1/2$ as
long as the market run. The market activity stops once all the expectations
fall inside the interval whose width is smaller than $2c$. This happens
after time $\widehat{T}\sim -(1+\log _{2}c)/q_{\gamma }$.

\section{Simulation results}

When the price expectations of agents have both the adaptive component and
confirmatory bias, obtaining analytical results is beyond reach. We thus
proceed by running numerical simulations. In what follows, we vary the
values of $\alpha $ and $\sigma $, from $0$ to $1$, in steps of 0.01. For
each pair of values ($\alpha ,\sigma $) the market is simulated 10 times.
The cost of a trading transaction is fixed at $c=0.005$. Each simulation can
run for 100 steps, being this a time interval large enough for the market
price to converge to the steady-state. Note that in the simulations we
define the steady-state as the situation in which the difference between the
market prices in periods $t$ and $t+1$ differ by a value smaller than
0.0001. We then look at the degree of market informational inefficiency in
the long-run, i.e. how far the market price diverges from the fundamental
value of the asset (averaging across the 10 simulations). We also look at
the average volatility of the market price, as measured by the standard
deviation in the market price in the last 90\% of the steps of the
simulation.

The agents have a relatively low level of wealth. Remember that if the
outcomes of the trading strategy of a trader lead to losses that,
accumulated over several periods, exhaust her wealth, she quits the market
and is replaced by another trader with a randomly drawn initial price
expectation. Given that traders have a relatively low level of wealth, a
certain number of them will quit the market and this implies that the
turnover rate of traders is relatively high. This means that some amount of
noise gets continuously injected into the market.

\begin{center}
\textbf{[Insert Figure 5 about here]}
\end{center}

Figure 5 (Panels A, B, and C) report the informational inefficiency of the
market (as measured by the divergence of the steady-state market price from
the fundamental value of the asset) for the cases in which the fraction of
agents that revise their expectations in every period is $\gamma =0.2,$ $%
0.5, $ and $0.8$, respectively. The market inefficiency is a function of the
weight of the adaptive component in the price expectations of traders $%
\alpha $ and of the degree of confirmatory bias $\sigma $. Colors closer to
dark blue indicate lower level of market inefficiency, while those closer to
dark red indicate higher inefficiency. Figure 6 describes the volatility of
the market price, while Figure 7 shows the average number of traders that
exit the market as their wealth hits the zero bound.

\begin{center}
\textbf{[Insert Figures 6 and 7 about here]}
\end{center}

Analyzing these figures, we obtain the following findings.

Fixing the value of $\sigma $, as we move from the extreme-left point ($%
\alpha =0$) to the right, the average deviation of the long-run market price
from the fundamental value ($0.5$) first decreases and then increases - at
least for some values of $\sigma $. In other words:

\begin{proposition}
Market inefficiency can be non-monotonic in the weight of the adaptive
component ($\alpha $) in the price expectations of agents.
\end{proposition}

In all three figures, we find that for very high values of $\alpha $, the
degree of the \ informational inefficiency of the market is very high.
Clearly, when traders put a large weight to the past price in forming
expectations, the initial price becomes very important. When receiving
information which indicates that the value of the asset is low (even in the
absence of confirmatory bias), traders tend to give little weight to it -
basically, all that matters is the past price. In this case, the initial
price strongly influences the aggregate expectation formation process (the
expectations of all agents quickly converge upwards to some point between
the initial price and the fundamental value) and, given that in our case the
initial price strongly differs from the fundamental value, the long-run
market price stays largely above the fundamental value.

Consider now the situation with the most extreme form of confirmatory bias,
i.e. all traders completely ignore the information that comes from others.
As $\alpha $ declines, the traders give less weight to the past prices and
more weight to their own expectations of the previous period. Therefore, the
agents whose initial expectations are very low do not move their next-period
expectations upwards too much. At the same time, the market price keeps
falling, driven by the Walrasian auctioneer (which also implies the downward
move in the expectations of the agents whose initial expectations are high).
These two inter-related processes - the upward drift of price expectations
of initially low-expectation agents and the downward pressure on the market
price converge to some value relatively close to the fundamental one.

As $\alpha $ declines further, we observe that the market inefficiency rises
again. This is due to the fact that for the lower values of $\alpha $, the
first process (upward move in expectations of the initially low-expectation
agents) becomes slower than the second one (i.e. downward move in the market
price). Thus, the low-expectation agents keep making negative profits,
eventually hit the zero-wealth bound, and exit (we can note this by looking
at Figure 7: the number of agents that exit the market increases at the
bottom-left part of the figure). There is a sufficiently high exit rate of
these agents from the market so as to soften the downward move in the market
price, which means that the long-run price at which the system settles down
is higher than in the situation in which the exit of traders is negligible.

As the degree of confirmatory bias of agents becomes smaller (i.e. the value
of $\sigma $ increases), the channel that leads to the exit of
low-expectation traders softens down, as there is now an additional
mechanism that creates an upward pressure on the expectations of those
traders: the integration of information that comes from their peers. Notice
(in Figure 7) that the exit rate is lower at the higher values of $\sigma $.

Furthermore, comparing across the three panels of Figure 5, one notes that
as the fraction of agents that revise their expectations in each period ($%
\gamma $) increases, the areas of $\sigma $ in which the non-monotonicity in 
$\alpha $ occurs becomes smaller:

\begin{proposition}
The tendency of the market inefficiency to be non-monotonic in $\alpha $ is
stronger, the lower is the fraction of agents that revise their price
expectations in each period.
\end{proposition}

This happens because the higher frequency of expectation revision (larger $%
\gamma $) and the likelihood to integrate the information coming from other
traders (larger $\sigma $) act in a complementary fashion: if the rate of
revision of price expectations is relatively low, the "openness of mind"
(i.e. low degree of confirmatory bias) has a relatively small effect on the
mitigation of the exit channel. It is only when the agents revise their
expectations relatively frequently, that the "openness of mind" starts to
have a real bite, and the upward-sloping part on the left side of the
relation between market inefficiency and $\alpha $ starts to disappear.

Let's now fix the value of $\alpha =0.25$ on Figure 5A, $\alpha =0.1$ on
Figure 5B, and $\alpha =0.05$ on Figure 5C. As we move from the point at the
bottom ($\sigma =0$) upwards, the average deviation of the steady-state
market price from the fundamental value ($0.5$) first decreases and then
increases. In other words:

\begin{proposition}
Market inefficiency can be non-monotonic in the degree of confirmatory bias
of agents.
\end{proposition}

The first part of the non-monotonic relationship is easy to explain: as an
agent suffers less from confirmatory bias, she starts to integrate at least
some of the information about the fundamentals contained in the price
expectations of another trader (incidentally, this phenomenon occurs only
when the adaptive component in the price expectation is relatively small).

But why the market inefficiency would \textit{rise }as the agents become
even more "open-minded"? To understand this, we need to note that this
phenomenon occurs only when the adaptive component is not too small. Then,
the fact that the initial price differs substantially from the fundamental
value plays a role. The agents have an early-stage upward drift in
expectations. At the same time, the market price starts to fall. If the
agents are very "open-minded", this implies that they 'excessively'
integrate the early upward price drift into their expectations, which, in
turn, implies that the price at which the market settles in the long run is
relatively high. If, instead, the agents' confirmatory bias is stronger, the
decline in the market price is faster than the 'propagation' of the
upward-drifting expectations: this is why the market settles at the price
relatively close to the fundamental value.

This analysis suggests a very interesting and potentially more general
insight: when market participants suffer from more than one deviation from
fully rational behavior (in our case, adaptive expectations and confirmatory
bias), at least in some range of bias parameters, the two biases mitigate
each other. Given our analysis, it should not be difficult to construct
examples of asset markets with traders that have multiple sources of biases,
that exhibit the same price behavior as under full rationality, and in which
the price behavior would deviate from the full-rationality benchmark as soon
as one of the bias sources is eliminated.

Next, looking across the three panels of Figure 5, we also note that the
values of $\alpha $ at which we have found the non-monotonicity of the
market inefficiency decrease at higher values of $\gamma $. In other words:

\begin{proposition}
The weight of the adaptive component in the expectations ($\alpha $) at
which the non-monotonicity of the market inefficiency in the degree of
confirmatory bias occurs decreases with the fraction of agents that revise
their expectations in every period.
\end{proposition}

To capture the intuition behind this result, we need to conduct the
following thought experiment. Let's fix a point with sufficiently high
values of $\alpha $ and $\sigma $, for example $(0.4,0.4)$. Next, let's
increase the frequency of revision of expectations by the agents, from $%
\gamma =0.2$ to $\gamma =0.8$. We then observe that the market inefficiency
increases.\ This indicates the importance of the frequency with which agents
revise their price expectations for the propagation of the 'excessive'
integration of the upward drift into the expectations, as noted above. In
other words, at higher frequency of expectation revision, this 'excessive'
integration of the upward drift channel swamps the opposite (i.e. the
quantity-of-information) channel more easily. The quantity-of-information
channel starts to play a role only when the early-stage upward drift is
sufficiently small (i.e., history weighs relatively little in the
expectation formation).

If we measure the degree of market inefficiency, while varying $\sigma $
along a fixed $\alpha $, in ranges different from those where the
non-monotonicity occurs (for example, $\alpha =0.1$ and $\alpha =0.3$ on
Figure 5A), we see that at the higher values of $\alpha $ the relationship
between the degree of market inefficiency and $\sigma $ is negative, while
at the lower values of $\alpha $, this relationship is positive. In other
words:

\begin{proposition}
The slope of the relationship of market inefficiency in the degree of
confirmatory bias ($\sigma $) can be of opposite sign at different values of
the weight of the adaptive component ($\alpha $).
\end{proposition}

The above discussion has already hinted at the potential explanation why
this reversal of the relationship occurs. At sufficiently high values of $%
\alpha $, the early-stage upward drift is very important and the smaller
confirmatory bias of agents only helps to propagate this drift into the
price expectations. At sufficiently low values of $\alpha $, the early-stage
upward drift matters much less and the smaller confirmatory bias becomes
beneficial for the informational efficiency of the market, because it helps
to integrate more of the relatively unbiased information into the
expectations. In other words, in both cases the smaller confirmatory bias
(i.e. higher $\sigma $) plays the role of the catalyzer; what differs in the
two cases is the initial unbiasedness of expectations.

\section{Conclusion}

This paper has studied the informational efficiency of an agent-based
financial market with a single traded asset. The price initially differs
from the fundamental value, about which the agents have noisy private
information (which is, on average, correct). A fraction of traders revise
their price expectations in each period. The price at which the asset is
traded is public information. The agents' expectations have an adaptive
component (i.e. the past price influences their future price expectation to
some extent) and a social-interactions component with confirmatory bias
(i.e. agents exchange information with their peers and tend to discard the
information that differs too much from their priors).

We find that the degree of informational inefficiency of the market
(measured as the deviation of the long-run market price from the fundamental
value of the asset) can be non-monotonic both in the weight of the adaptive
component and in the degree of the confirmatory bias. For some ranges of
parameters, two biases tend to mitigate each other's effect, thus increasing
the informational efficiency.

Our findings complement the well-known results in the theory of markets
showing that the allocative efficiency can be obtained even under
substantial deviation from individual rationality of agents (Gode and Sunder
1993, 1997). We show that deviations from individual rationality, under
certain conditions, can also facilitate the informational efficiency of
markets. The key condition for this property is that the various behavioral
biases that agents possess should mutually dampen their effects on the price
dynamics.

Given the potential importance of this insight for financial economics, the
natural extension of this work is to test its' predictions experimentally.
This would require to construct experimental financial markets with human
traders, similar to the setting of Haruvy et al. (2007), with the additional
feature of allowing agents to share their information (in some restricted
form). The outcomes of interest in such an experiment would be both the
evolution of market price of the asset and the elicited price expectations
of traders.

 \begin{figure}
 \centering
 \includegraphics[width=7cm]{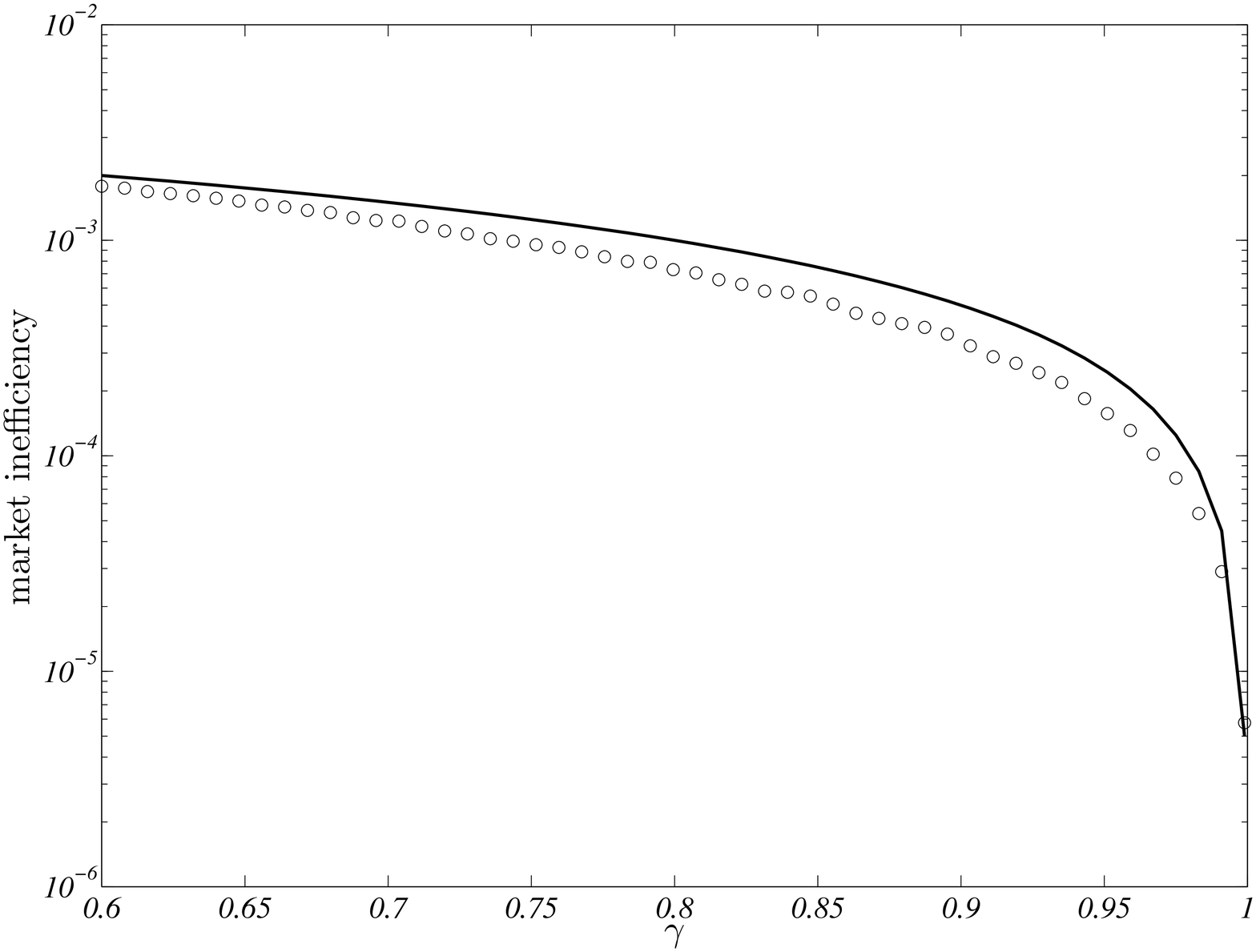}  \quad
 \includegraphics[width=7cm]{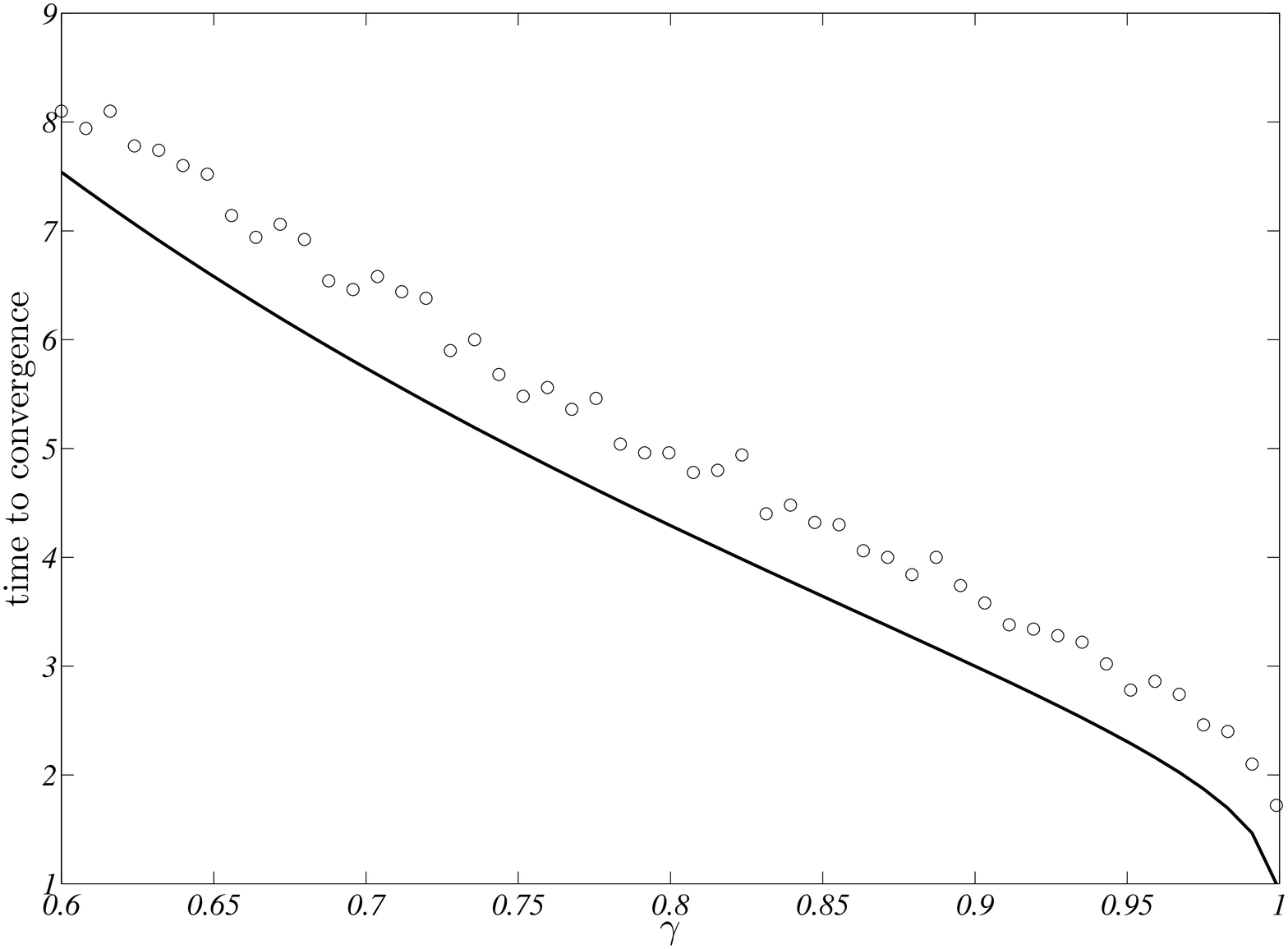}  
\caption{Purely adaptive expectations ($\alpha=1$). Left panel: semilog plot
  of the market inefficiency as a 
  function of $\gamma$. Right panel: time to convergence to the steady state
  as a function of 
  $\gamma$. There are $N=1000$ agents, the transaction cost $c=0.005$. Each
  point represents the average over $50$ simulations. $\bigcirc$ : numerical  
  simulations, solid line : analytical results.}   
 \label{fig:alpha1}
 \end{figure}

 \begin{figure}
 \centering
 \includegraphics[width=7cm]{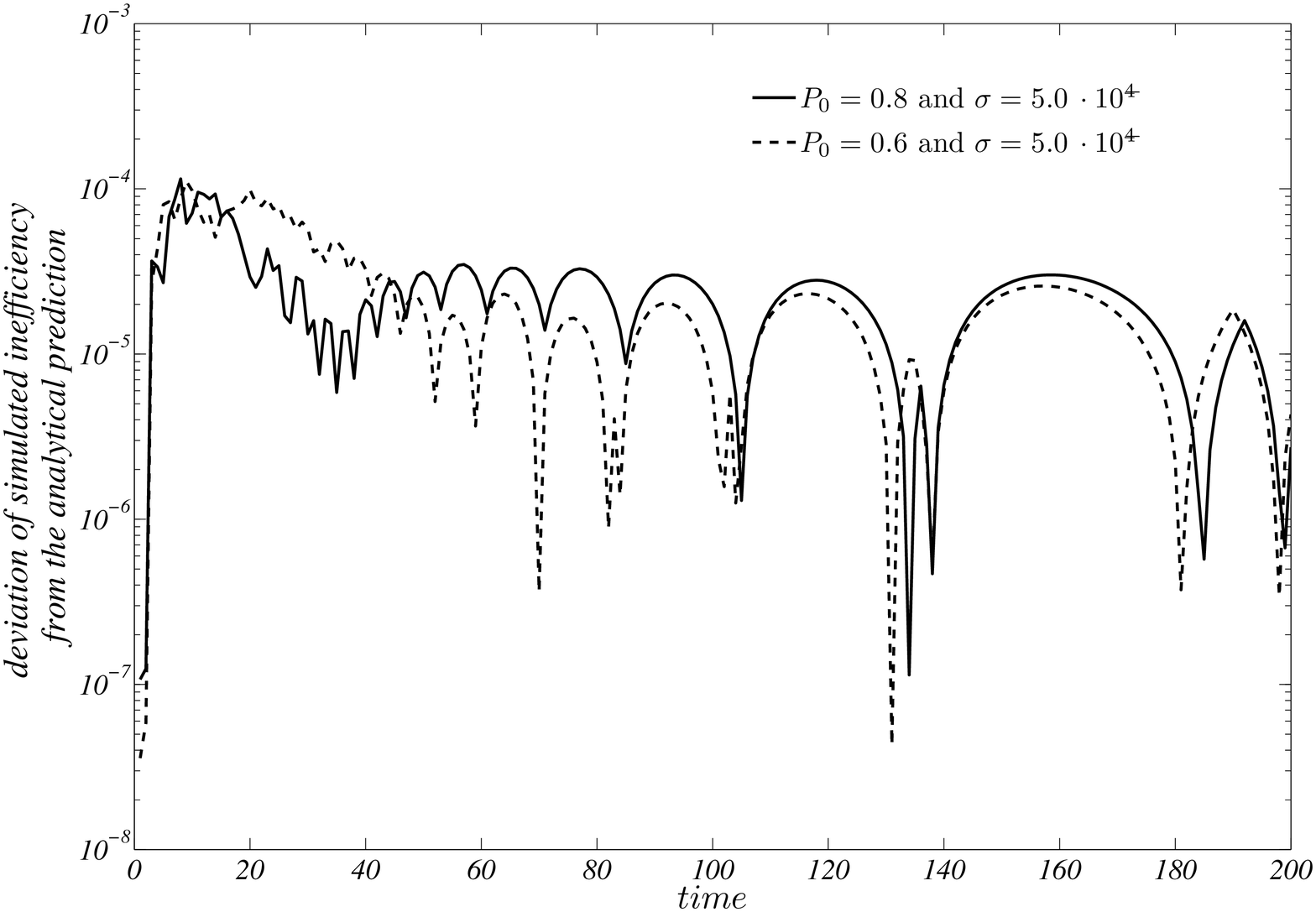}  \quad
 \includegraphics[width=7cm]{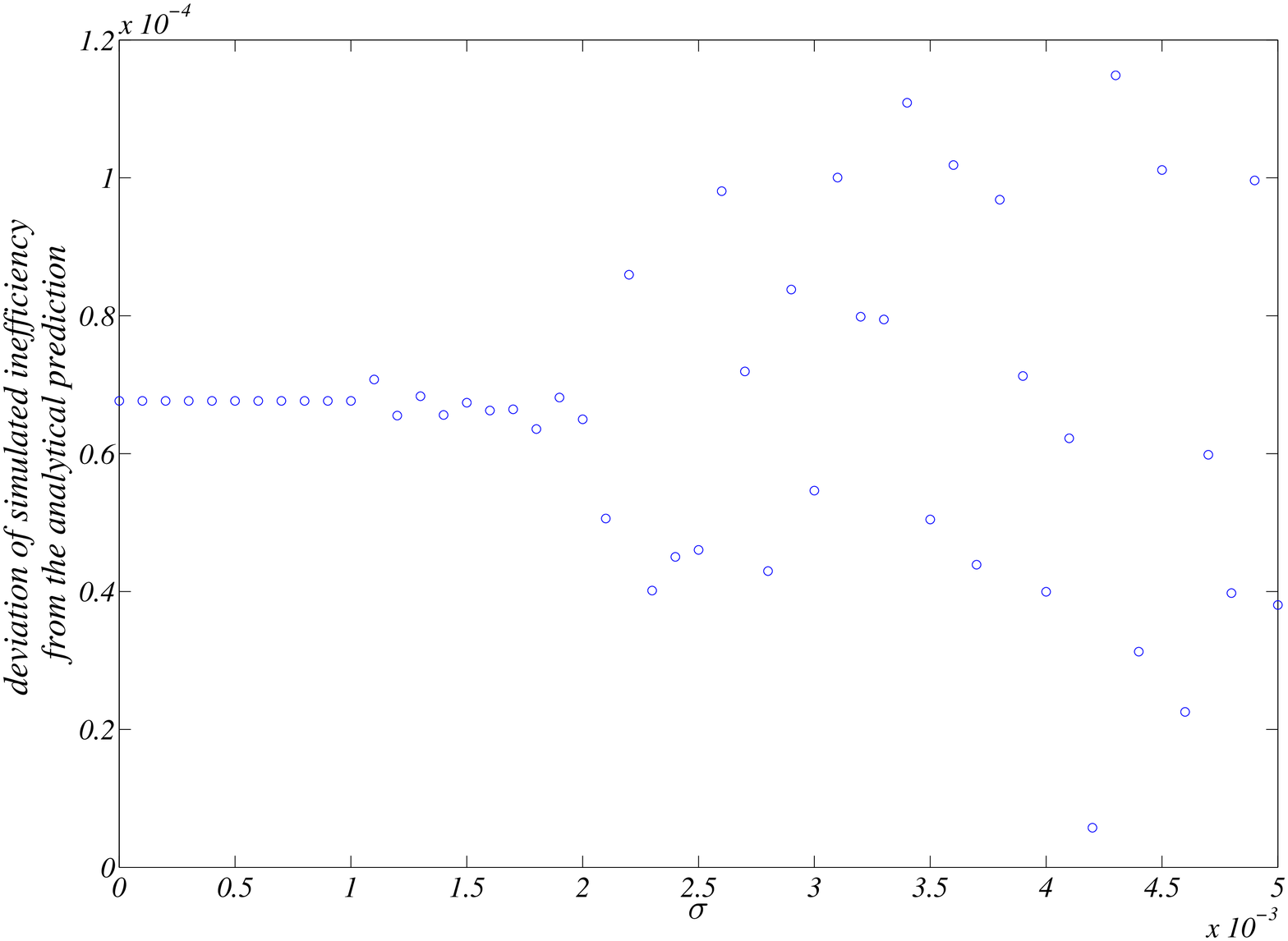}  
\caption{Social interaction without adaptive component ($\alpha=0$ and $\sigma
  \sim 0$). Left panel: semilog 
  plot of the  difference between the simulated and analytical market
  inefficiency with $\sigma=5.0\, \cdot
  10^{-4}$ and $P_0=0.6$ or $0.8$. Right pane:l difference between the
  simulated and analytical market  
  inefficiency  as a function of $\sigma$. There are $N=1000$
  agents, 
  the transaction cost $c=0.005$.}    
 \label{fig:alpha0sigmasmall}
 \end{figure}

 \begin{figure}
 \centering
 \includegraphics[width=7cm]{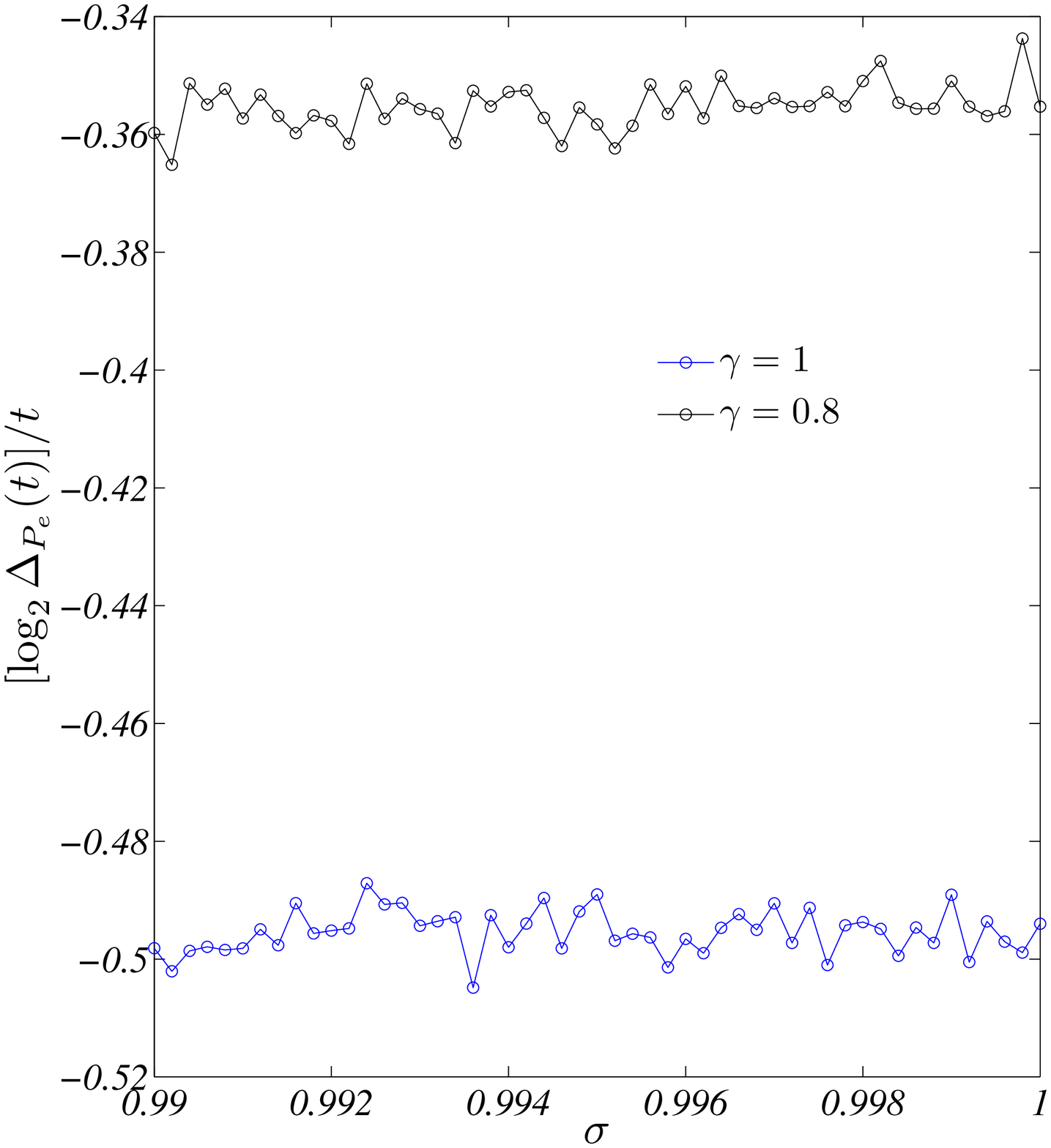}
\quad
 \includegraphics[width=7cm]{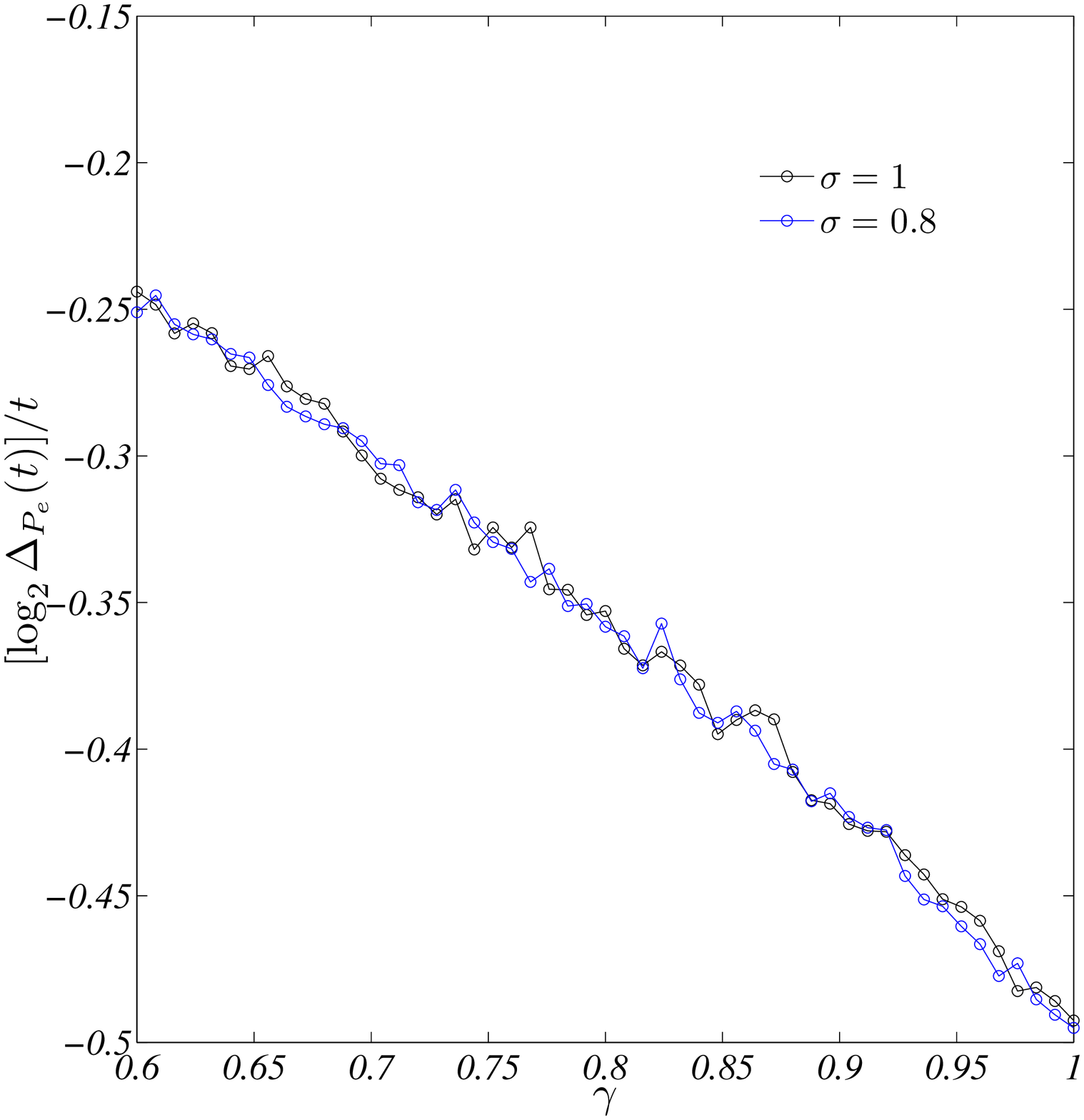}
\caption{Evolution of $[\log_2\Delta_{P_e}(t)]/t$ as a function of
  $\sigma$ (left panel) and as a function of $\gamma$ (right panel). There are
  $N=1000$ agents, the transaction cost $c=0.005$.}    
 \label{fig:deltago20}
 \end{figure}

 \begin{figure}
 \centering
 \includegraphics[width=7cm]{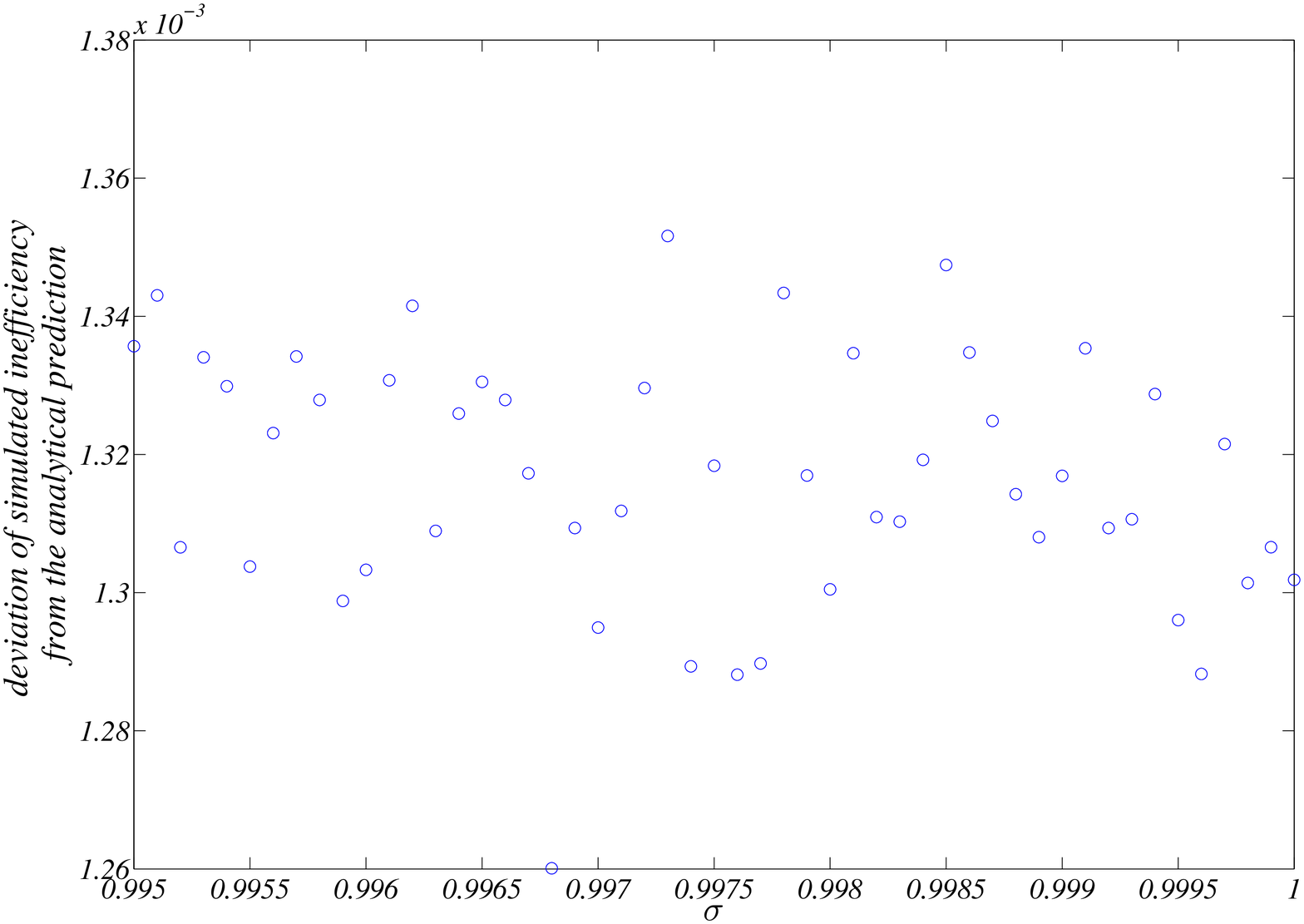}  \quad
 \includegraphics[width=7cm]{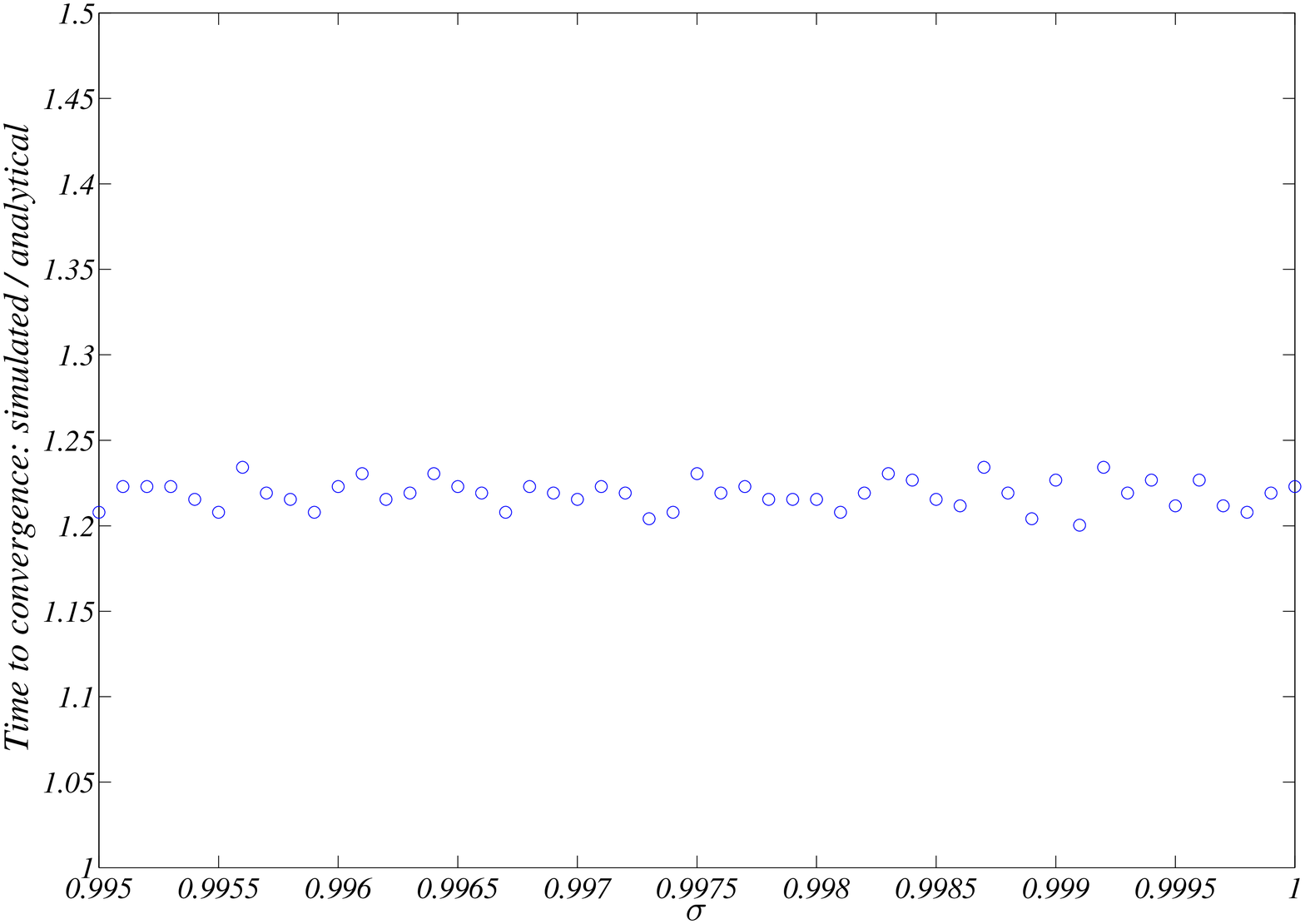}  
\caption{Social interaction without adaptive component ($\alpha=0$ and $\sigma
  \sim 0$). Left panel: difference between the simulated and analytical market
  inefficiency as a function of $\sigma$. Right
  panel: the ration of the simulated time to convergence over the analytical
  one, $\sigma$. There are
  $N=1000$ agents, the transaction cost $c=0.005$.}    
 \label{fig:alpha0sigmalarge}
 \end{figure}

\begin{figure}[htbp]
\centering
\includegraphics[width=5.5cm]{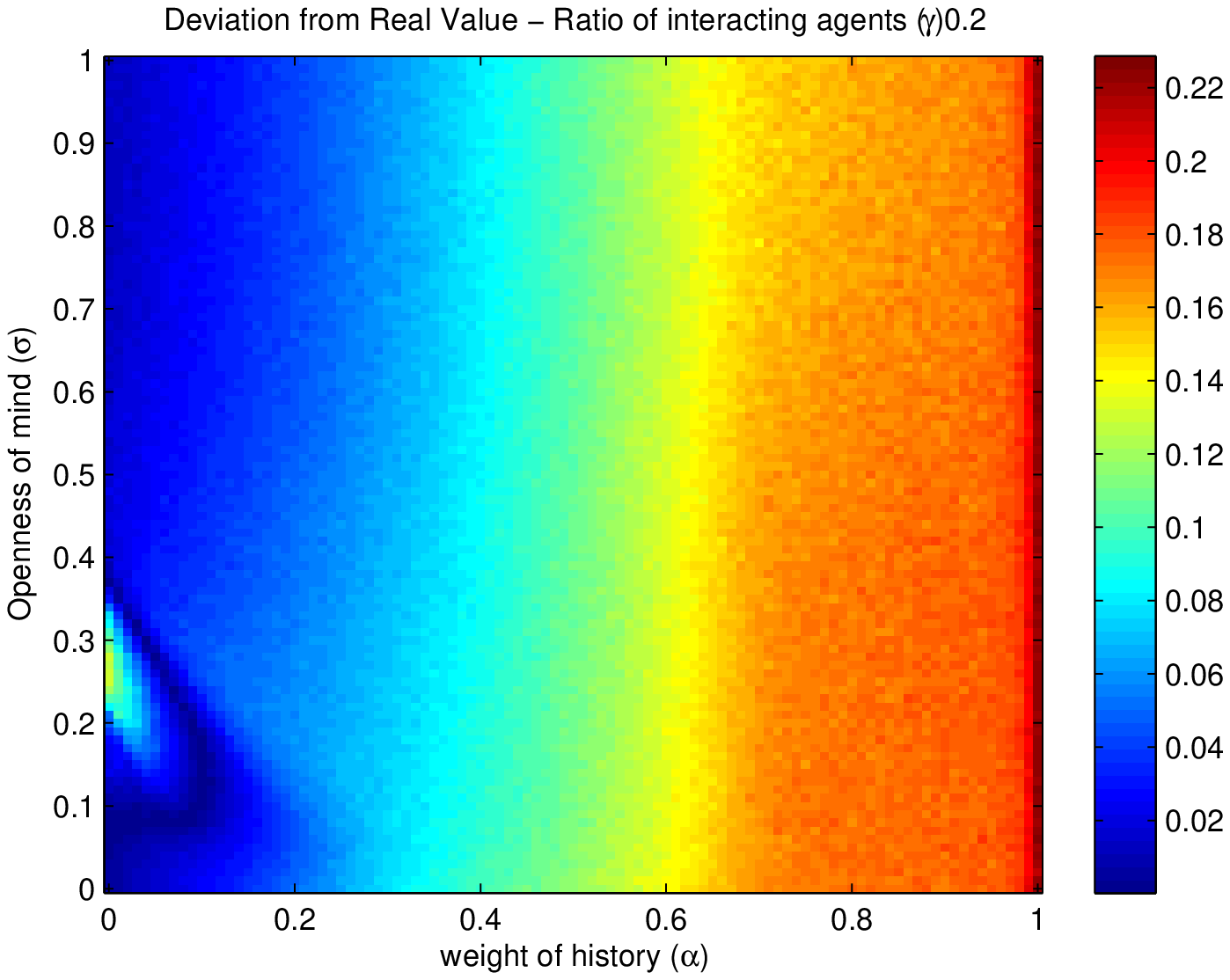}  
\includegraphics[width=5.5cm]{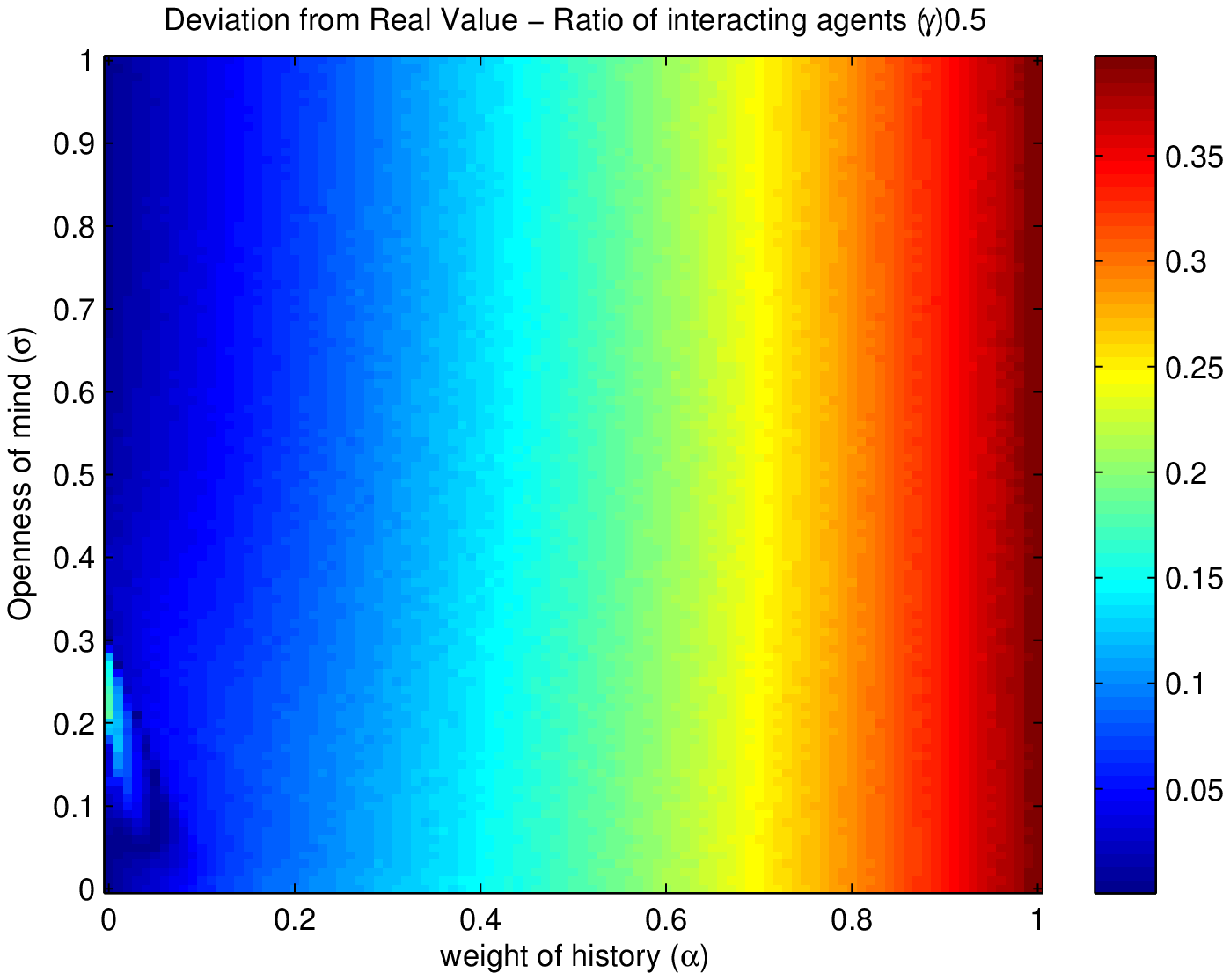}
\includegraphics[width=5.5cm]{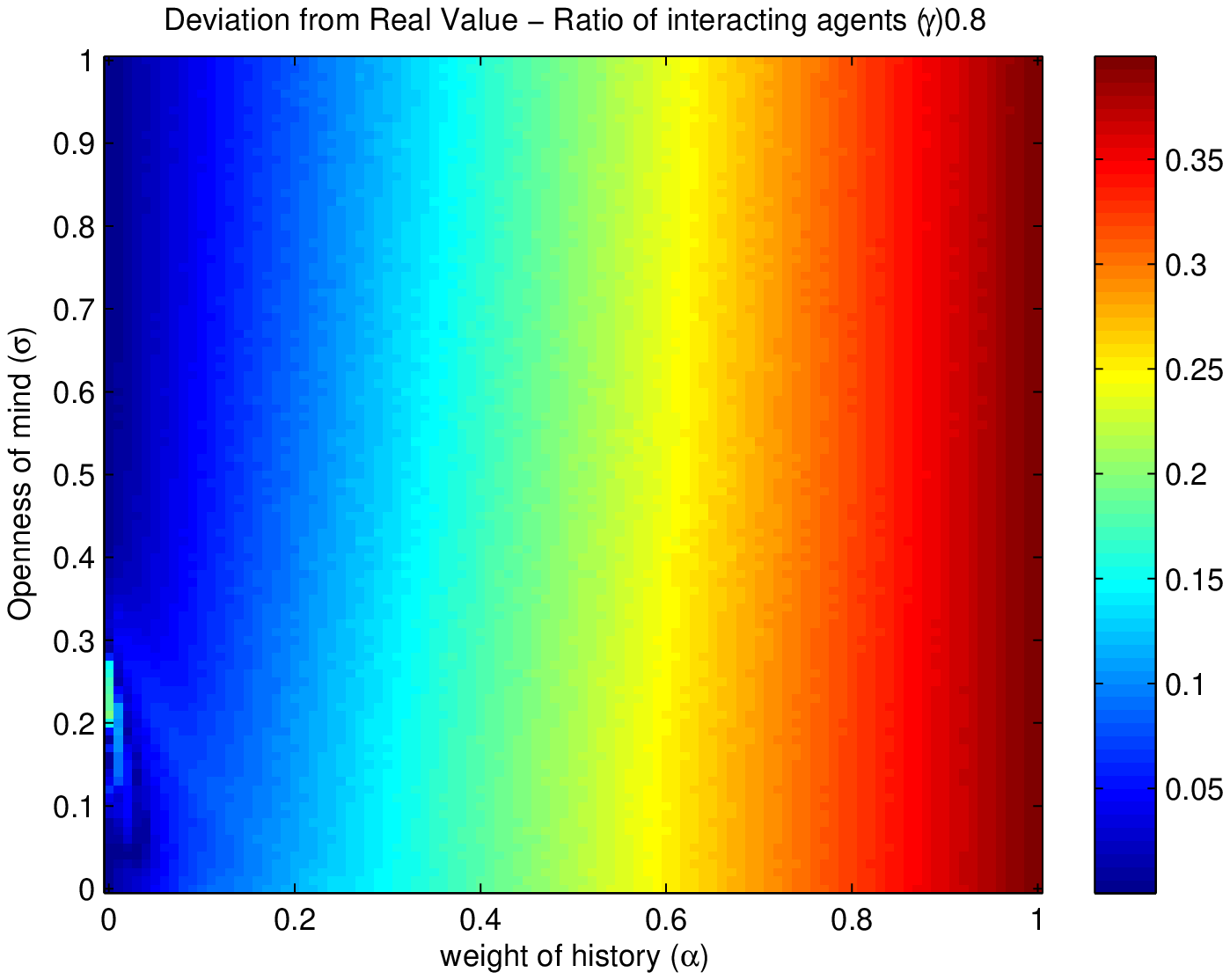}
\caption{The degree of market efficiency as a function of the adaptive
  component ($\alpha$) and the confirmatory bias ($\sigma$). Panel A:
  $\gamma=0.2$, Panel B $\gamma=0.5$ and Panel C : $\gamma=0.8$. The initial
  price is $0.9$.}  
\label{fig:SmallLW_scostamentoAvsSIGMAforGAMMAvar}
\end{figure}

\begin{figure}[htbp]
\centering
\includegraphics[width=5.5cm]{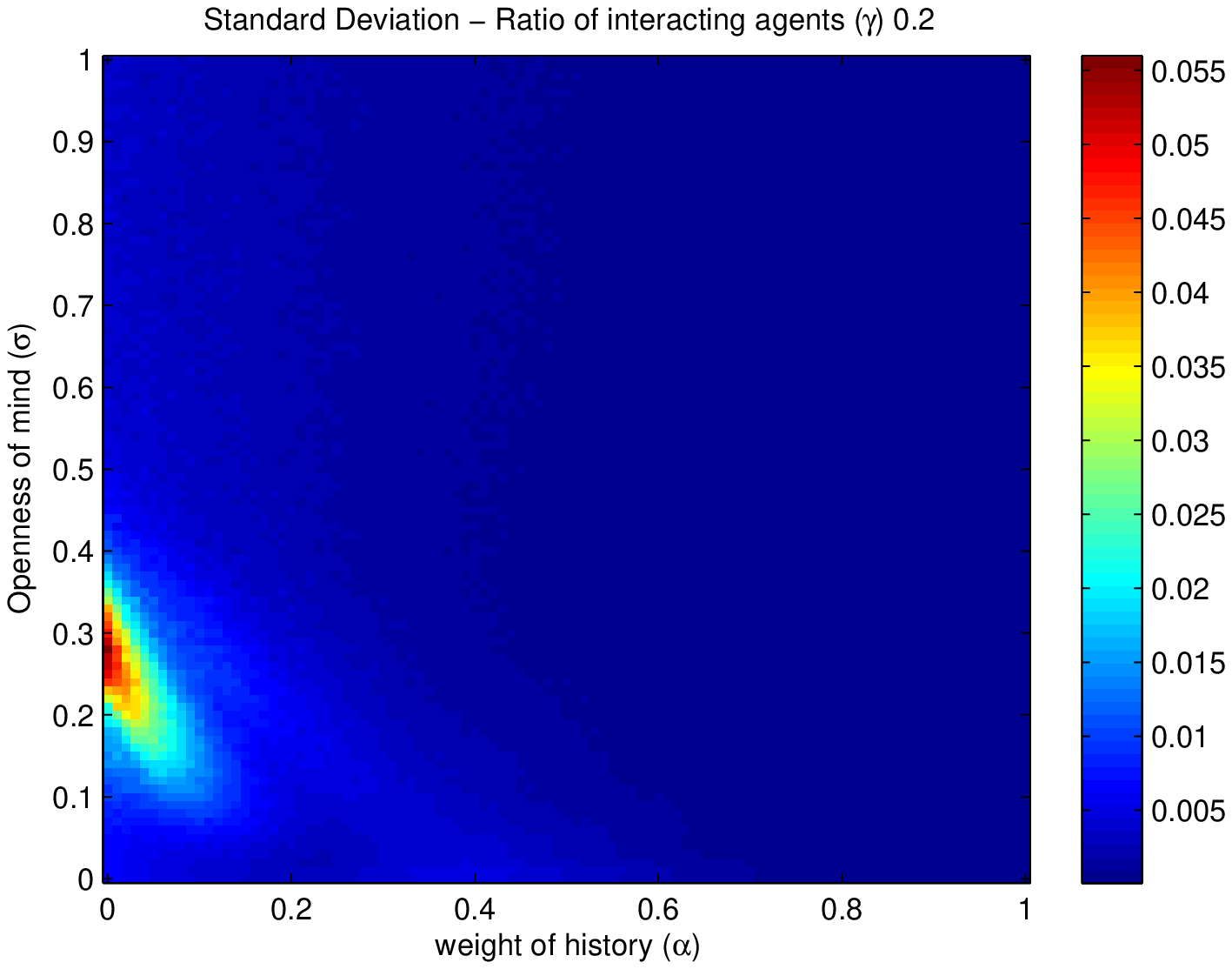}  
\includegraphics[width=5.5cm]{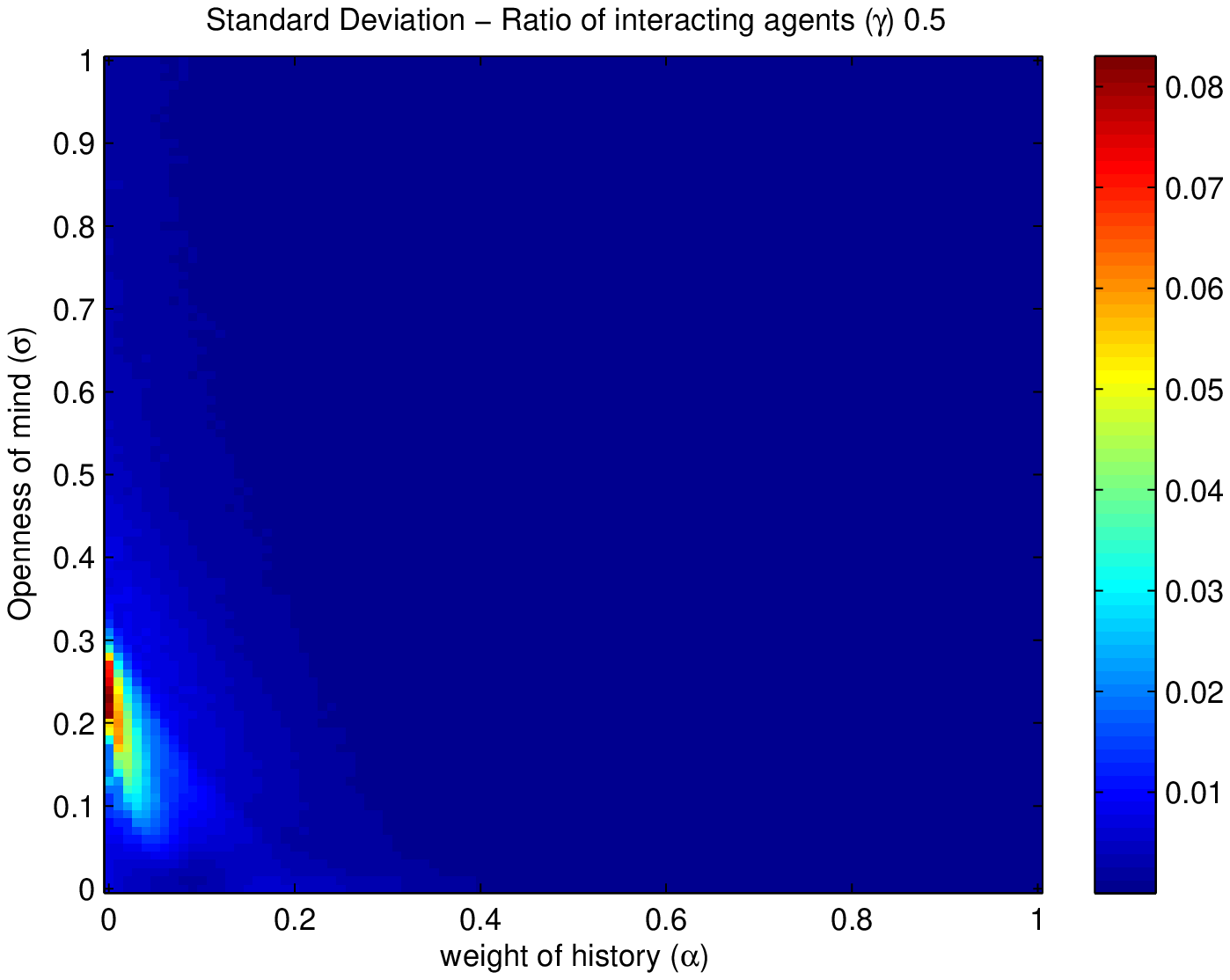}  
\includegraphics[width=5.5cm]{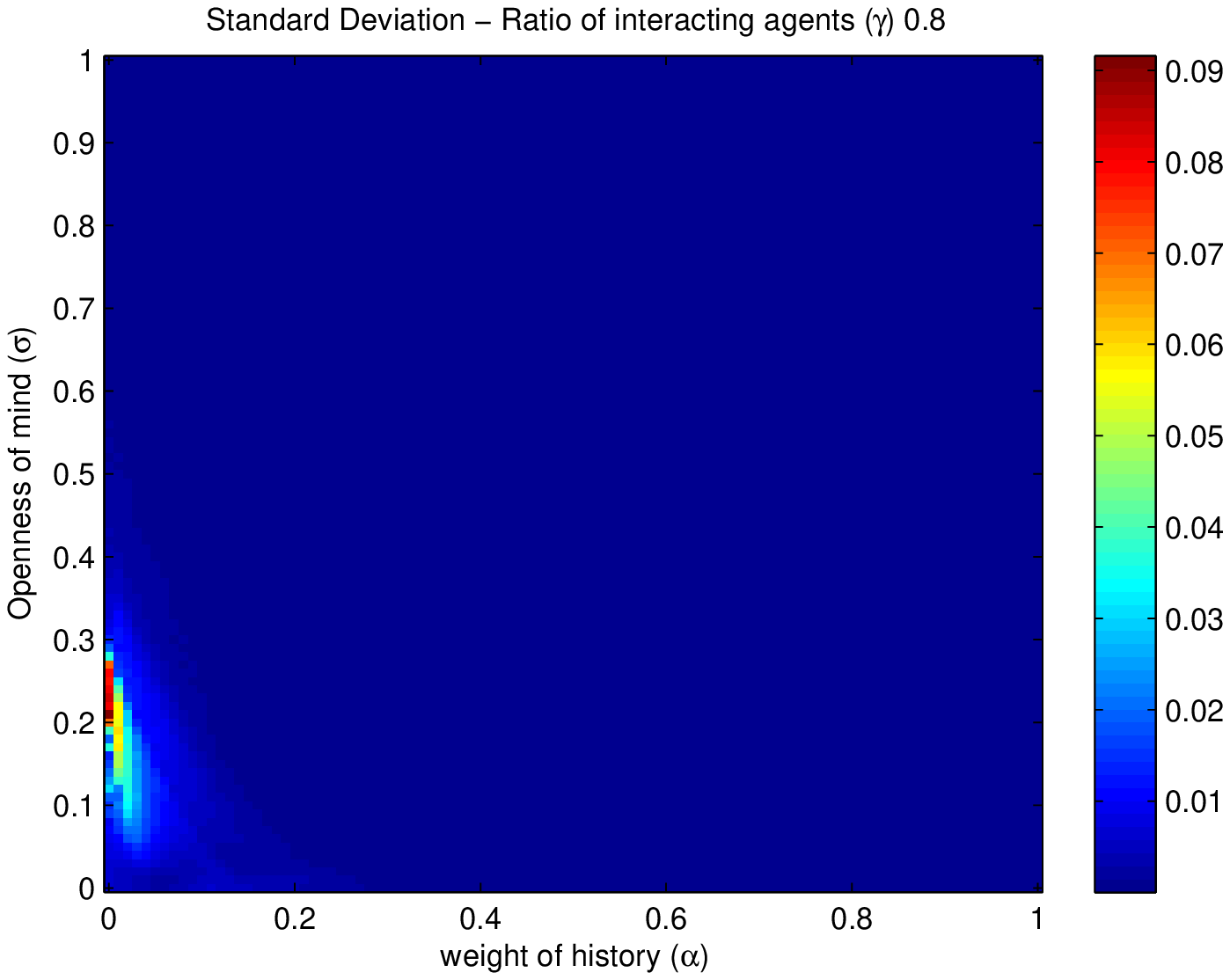}  
\caption{Volatility of the market price (calculated over the last $90
  \%$ of the time span) as a function of the adaptive
  component ($\alpha$) and the confirmatory bias ($\sigma$). Panel A:
  $\gamma=0.2$, Panel B $\gamma=0.5$ and Panel C : $\gamma=0.8$. The initial
  price is $0.9$.}  
\label{fig:SmallLW_devstdAvsSIGMAforGAMMAvar}
\end{figure}

\begin{figure}[htbp]
\centering
\includegraphics[width=5.5cm]{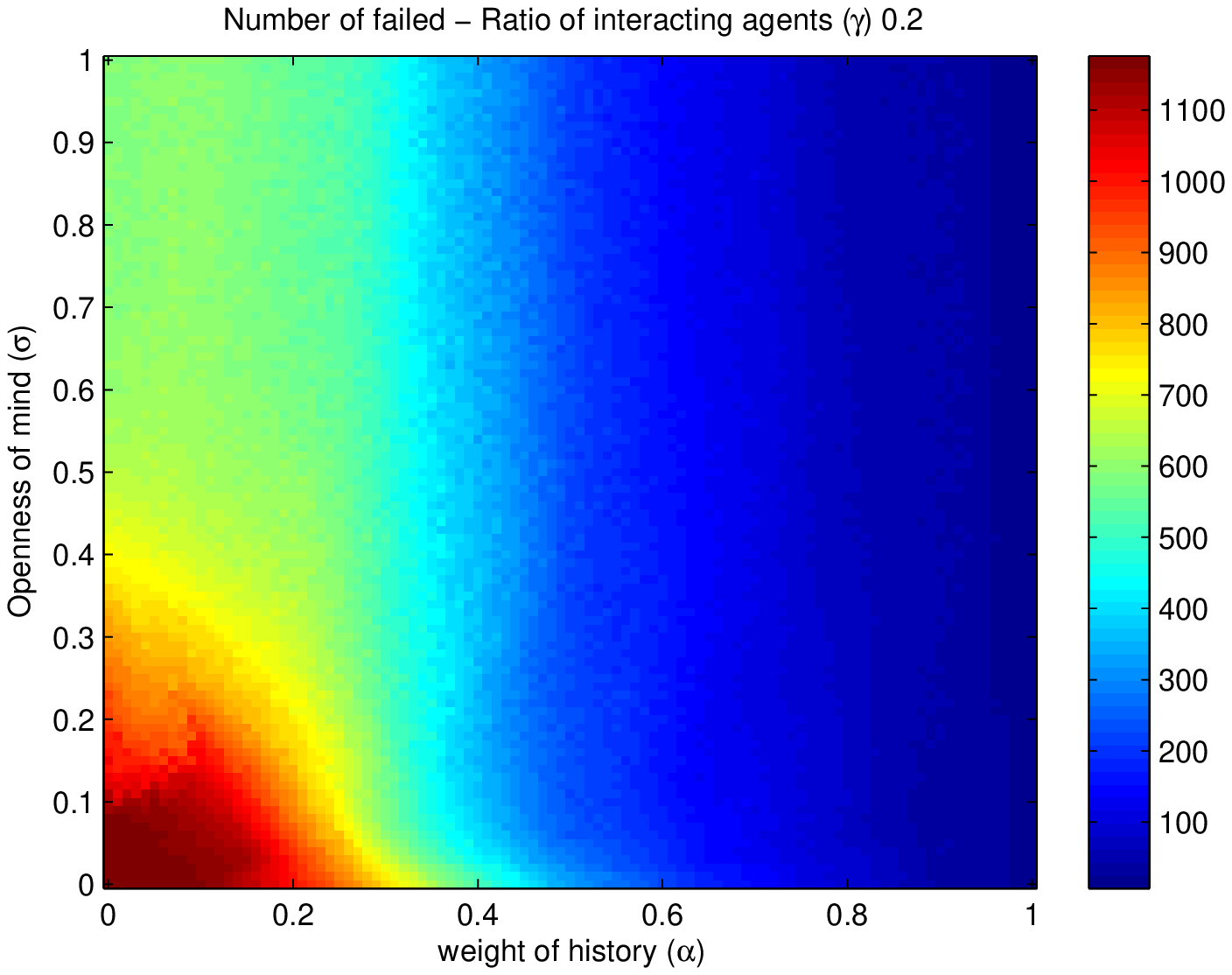} 
\includegraphics[width=5.5cm]{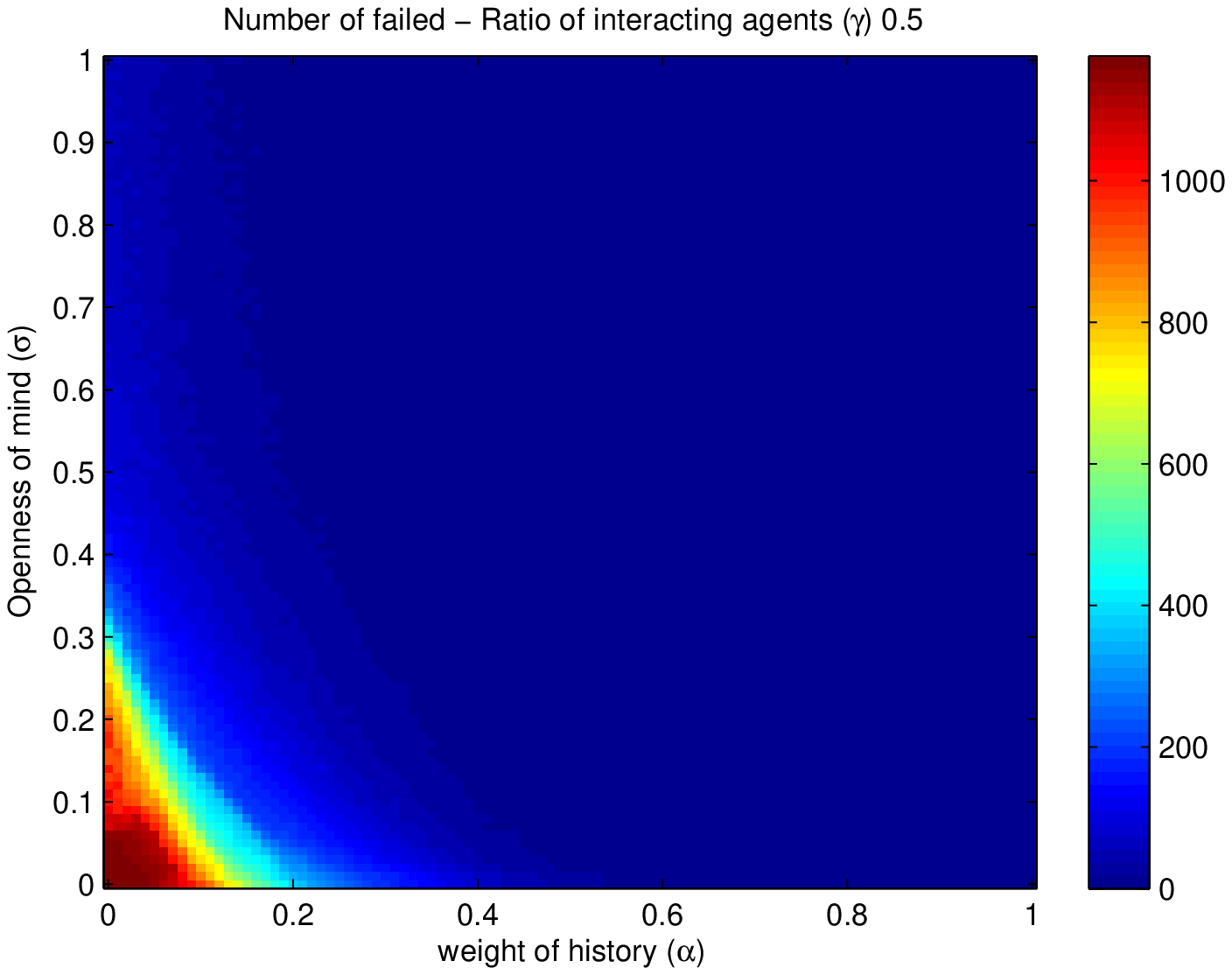}  
\includegraphics[width=5.5cm]{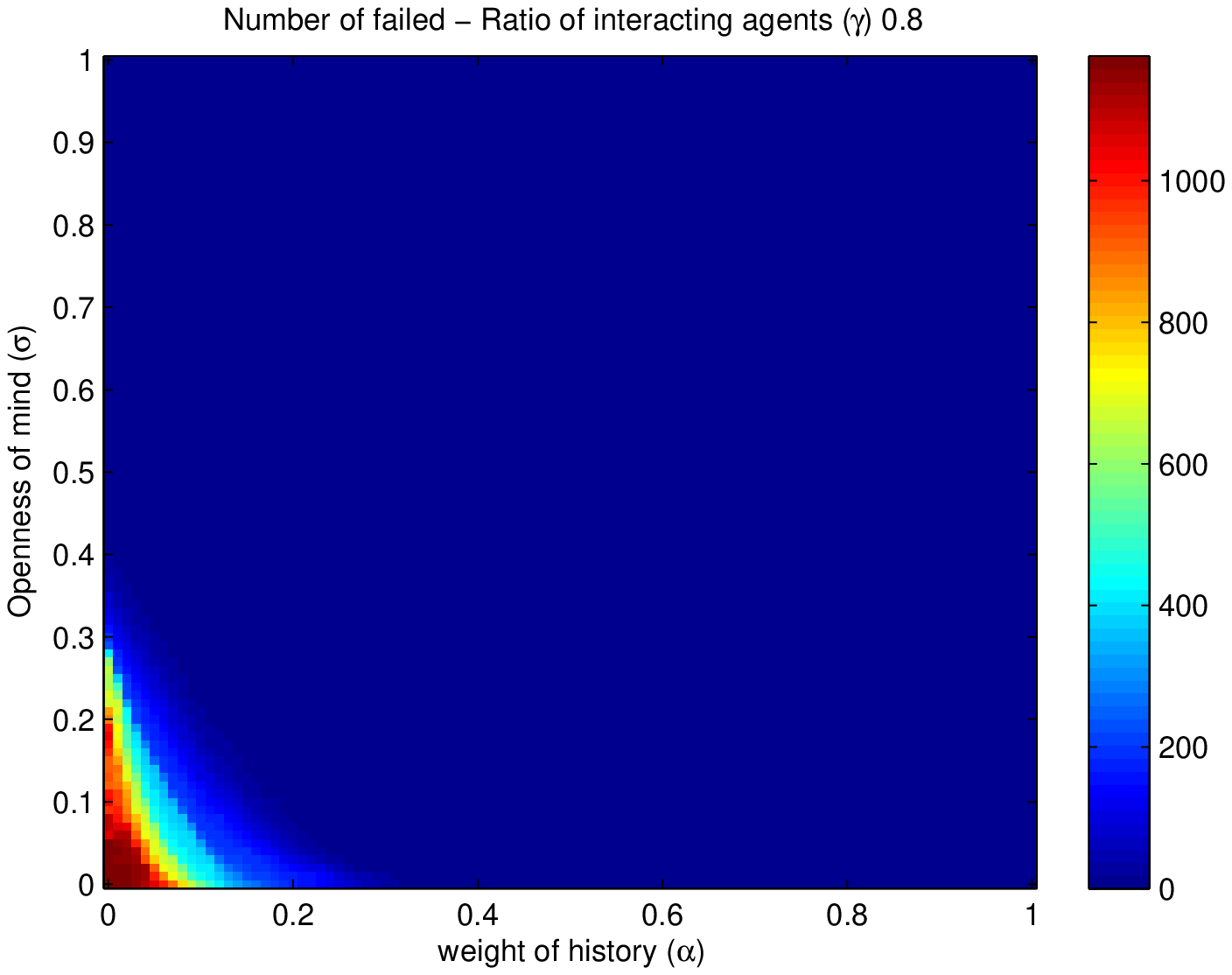}
\caption{Number of agents exiting the market  as a function of the adaptive
  component ($\alpha$) and the confirmatory bias ($\sigma$). Panel A:
  $\gamma=0.2$, Panel B $\gamma=0.5$ and Panel C : $\gamma=0.8$. The initial
  price is $0.9$.}  
\label{fig:SmallLW_NumMortiAvsSIGMAforGAMMAvar}
\end{figure}

\end{document}